\overfullrule=0pt
\input harvmac

\def\ap{{\alpha^{\prime}}}

\def\a{{\alpha}}

\def\l{{\lambda}}
\def\lb{{\overline\lambda}}
\def\wb{{\overline w}}
\def\b{{\beta}}

\def\g{{\gamma}}

\def\d{{\delta}}
\def\e{{\epsilon}}
\def\s{{\sigma}}
\def\k{{\kappa}}

\def\half{{1\over 2}}
\def\p{{\partial}}
\def\pb{{\overline\partial}}
\def\t{{\theta}}

\def\bar{\overline}
\def\({\left(}
\def\){\right)}
\def\cF{{\cal F}}

%\draft

\Title{\vbox{\hbox{AEI-2010-034}}} {\vbox{
    \centerline{\bf The Overall Coefficient of the Two-loop Superstring Amplitude}
    \centerline{\bf Using Pure Spinors}
 }
}

\bigskip\centerline{Humberto Gomez\foot{email: humgomzu@ift.unesp.br}$^{,a}$
and Carlos R. Mafra\foot{email: crmafra@aei.mpg.de}$^{,b}$}
\vskip .23in
\centerline{\it 
$^a$ Instituto de F\'\i sica Te\'orica
UNESP - Universidade Estadual Paulista}
\centerline{\it Caixa Postal 70532-2
01156-970 S\~ao Paulo, SP, Brazil}
\centerline{\it }
%\smallskip
\centerline{\it $^b$Max-Planck-Institut f\"ur
Gravitationsphysik, Albert-Einstein-Institut}
\smallskip
\centerline{\it 14476 Golm, Germany}
\medskip

\vskip .3in

Using the results recently obtained for computing integrals over (non-minimal) 
pure spinor superspace, we compute the coefficient of the massless two-loop 
four-point amplitude from first principles.
Contrasting with the mathematical difficulties in the RNS formalism where 
unknown normalizations of chiral determinant formul{\ae} force
the two-loop coefficient to be determined only indirectly through factorization,
the computation in the pure spinor formalism can be smoothly carried out.

\vskip .3in

\Date {March 2010}

\lref\superpoincare{
  N.~Berkovits,
  ``Super-Poincare covariant quantization of the superstring,''
  JHEP {\bf 0004}, 018 (2000)
  [arXiv:hep-th/0001035].
  %%CITATION = JHEPA,0004,018;%%
}
\lref\multiloop{
  N.~Berkovits,
  ``Multiloop amplitudes and vanishing theorems using the pure spinor
  formalism for the superstring,''
  JHEP {\bf 0409}, 047 (2004)
  [arXiv:hep-th/0406055].
  %%CITATION = JHEPA,0409,047;%%
}
\lref\twoloop{
  N.~Berkovits,
  ``Super-Poincare covariant two-loop superstring amplitudes,''
  JHEP {\bf 0601}, 005 (2006)
  [arXiv:hep-th/0503197].
  %%CITATION = JHEPA,0601,005;%%
}
\lref\twolooptwo{
  N.~Berkovits and C.~R.~Mafra,
  ``Equivalence of two-loop superstring amplitudes in the pure spinor and  RNS
  formalisms,''
  Phys.\ Rev.\ Lett.\  {\bf 96}, 011602 (2006)
  [arXiv:hep-th/0509234].
  %%CITATION = PRLTA,96,011602;%%
}
\lref\nekrasov{
  N.~Berkovits and N.~Nekrasov,
  ``Multiloop superstring amplitudes from non-minimal pure spinor formalism,''
  JHEP {\bf 0612}, 029 (2006)
  [arXiv:hep-th/0609012].
  %%CITATION = JHEPA,0612,029;%%
}
\lref\yuri{
  Y.~Aisaka and N.~Berkovits,
  ``Pure Spinor Vertex Operators in Siegel Gauge and Loop Amplitude
  Regularization,''
  JHEP {\bf 0907}, 062 (2009)
  [arXiv:0903.3443 [hep-th]].
  %%CITATION = JHEPA,0907,062;%%
}
\lref\grassi{
  P.~A.~Grassi and P.~Vanhove,
  ``Higher-loop amplitudes in the non-minimal pure spinor formalism,''
  JHEP {\bf 0905}, 089 (2009)
  [arXiv:0903.3903 [hep-th]].
  %%CITATION = JHEPA,0905,089;%%
}
\lref\PSSuperspace{
  N.~Berkovits,
  ``Explaining pure spinor superspace,''
  arXiv:hep-th/0612021.
  %%CITATION = HEP-TH/0612021;%%
}
\lref\nonmin{
  N.~Berkovits,
  ``Pure spinor formalism as an N = 2 topological string,''
  JHEP {\bf 0510}, 089 (2005)
  [arXiv:hep-th/0509120].
  %%CITATION = JHEPA,0510,089;%%
}
\lref\humberto{
  H.~Gomez,
  ``One-loop Superstring Amplitude From Integrals on Pure Spinors Space,''
  JHEP {\bf 0912}, 034 (2009)
  [arXiv:0910.3405 [hep-th]].
  %%CITATION = JHEPA,0912,034;%%
}
\lref\anomalia{
  N.~Berkovits and C.~R.~Mafra,
  ``Some superstring amplitude computations with the non-minimal pure spinor
  formalism,''
  JHEP {\bf 0611}, 079 (2006)
  [arXiv:hep-th/0607187].
  %%CITATION = JHEPA,0611,079;%%
}
\lref\mafraids{
  C.~R.~Mafra,
  ``Pure Spinor Superspace Identities for Massless Four-point Kinematic
  Factors,''
  JHEP {\bf 0804}, 093 (2008)
  [arXiv:0801.0580 [hep-th]].
  %%CITATION = JHEPA,0804,093;%%
}
\lref\dhokerVI{
  E.~D'Hoker and D.~H.~Phong,
  ``Two-Loop Superstrings VI: Non-Renormalization Theorems and the 4-Point
  Function,''
  Nucl.\ Phys.\  B {\bf 715}, 3 (2005)
  [arXiv:hep-th/0501197].
  %%CITATION = NUPHA,B715,3;%%
}
\lref\bigHowe{
  P.~S.~Howe,
  ``Pure Spinors Lines In Superspace And Ten-Dimensional Supersymmetric
  Theories,''
  Phys.\ Lett.\  B {\bf 258}, 141 (1991)
  [Addendum-ibid.\  B {\bf 259}, 511 (1991)].
  %%CITATION = PHLTA,B258,141;%%
\semi
  P.~S.~Howe,
  ``Pure Spinors, Function Superspaces And Supergravity Theories In
  Ten-Dimensions And Eleven-Dimensions,''
  Phys.\ Lett.\  B {\bf 273}, 90 (1991).
  %%CITATION = PHLTA,B273,90;%%
}
\lref\dhokerS{
  E.~D'Hoker, M.~Gutperle and D.~H.~Phong,
  ``Two-loop superstrings and S-duality,''
  Nucl.\ Phys.\  B {\bf 722}, 81 (2005)
  [arXiv:hep-th/0503180].
  %%CITATION = NUPHA,B722,81;%%
}
\lref\teseM{
  C.~R.~Mafra,
  ``Superstring Scattering Amplitudes with the Pure Spinor Formalism,''
  arXiv:0902.1552 [hep-th].
  %%CITATION = ARXIV:0902.1552;%%
}

\lref\Beilinson{
  A.~A.~Beilinson and Yu.~I.~Manin,
  ``The Mumford Form and the Polyakov Measure in String Theory,''
  Commun.\ Math.\ Phys.\  {\bf 107}, 359 (1986).
  %%CITATION = CMPHA,107,359;%%
}

\lref\dhokerRev{
  E.~D'Hoker and D.~H.~Phong,
  ``The Geometry of String Perturbation Theory,''
  Rev.\ Mod.\ Phys.\  {\bf 60}, 917 (1988).
  %%CITATION = RMPHA,60,917;%%
}

\lref\multiloopDHoker{
  E.~D'Hoker and D.~H.~Phong,
  ``Multiloop Amplitudes For The Bosonic Polyakov String,''
  Nucl.\ Phys.\  B {\bf 269}, 205 (1986).
  %%CITATION = NUPHA,B269,205;%%
}

\lref\Weisberger{
  W.~I.~Weisberger,
  ``Normalization of the Path Integral Measure and the Coupling Constants for
  Bosonic Strings,''
  Nucl.\ Phys.\  B {\bf 284}, 171 (1987).
  %%CITATION = NUPHA,B284,171;%%
}

\lref\polchinski{
  J.~Polchinski,
  ``String theory. Vol. 1: An introduction to the bosonic string,''
%\href{/spires/find/hep/www?irn=4634799}{SPIRES entry}
{\it  Cambridge, UK: Univ. Pr. (1998) 402 p}
}
\lref\PolchinskiFactor{
  J.~Polchinski,
  ``Factorization of Bosonic String Amplitudes,''
  Nucl.\ Phys.\  B {\bf 307}, 61 (1988).
  %%CITATION = NUPHA,B307,61;%%
}

\lref\farkas{
Hershel M.~Farkas and Irwin Kra, 
``Riemann Surfaces'', Second Edition, Springer-Verlag, New York, 1992.
}

\lref\bershadsky{
  M.~Bershadsky, S.~Cecotti, H.~Ooguri and C.~Vafa,
  ``Kodaira-Spencer theory of gravity and exact results for quantum string
  amplitudes,''
  Commun.\ Math.\ Phys.\  {\bf 165}, 311 (1994)
  [arXiv:hep-th/9309140].
  %%CITATION = CMPHA,165,311;%%
}

\lref\greenS{
  M.~B.~Green and M.~Gutperle,
  ``Effects of D-instantons,''
  Nucl.\ Phys.\  B {\bf 498}, 195 (1997)
  [arXiv:hep-th/9701093].
  %%CITATION = NUPHA,B498,195;%%
}

\lref\polchinskiII{
  J.~Polchinski,
  ``String theory. Vol. 2: Superstring theory and beyond,''
%\href{/spires/find/hep/www?irn=4634802}{SPIRES entry}
{\it  Cambridge, UK: Univ. Pr. (1998) 531 p}
}

\lref\FORM{
  J.~A.~M.~Vermaseren,
  ``New features of FORM,''
  arXiv:math-ph/0010025.
  %%CITATION = MATH-PH/0010025;%%
\semi
  M.~Tentyukov and J.~A.~M.~Vermaseren,
  ``The multithreaded version of FORM,''
  arXiv:hep-ph/0702279.
  %%CITATION = HEP-PH/0702279;%%
}

\lref\dhokerperturbation{
 E.~D'Hoker and D.~H.~Phong,
 ``The Geometry of String Perturbation Theory,''
  Rev.\ Mod.\ Phys.\  {\bf 60}, 917 (1988).
  %%CITATION = RMPHA,60,917;%%
}

\lref\weinbergU{
  S.~Weinberg,
  ``Coupling Constants And Vertex Functions In String Theories,''
  Phys.\ Lett.\  B {\bf 156}, 309 (1985).
  %%CITATION = PHLTA,B156,309;%%
\semi
  S.~Weinberg,
  ``Covariant Path Integral Approach To String Theory,''
  %%CITATION = C86/12/30;%%
}
\lref\bOda{
  I.~Oda and M.~Tonin,
  ``Y-formalism and $b$ ghost in the Non-minimal Pure Spinor Formalism of
  Superstrings,''
  Nucl.\ Phys.\  B {\bf 779}, 63 (2007)
  [arXiv:0704.1219 [hep-th]].
  %%CITATION = NUPHA,B779,63;%%
}

\lref\MedinaUm{
  R.~Medina, F.~T.~Brandt and F.~R.~Machado,
  ``The open superstring 5-point amplitude revisited,''
  JHEP {\bf 0207}, 071 (2002)
  [arXiv:hep-th/0208121].
  %%CITATION = JHEPA,0207,071;%%
\semi
  L.~A.~Barreiro and R.~Medina,
  ``5-field terms in the open superstring effective action,''
  JHEP {\bf 0503}, 055 (2005)
  [arXiv:hep-th/0503182].
  %%CITATION = JHEPA,0503,055;%%
}
\lref\stieOC{
  S.~Stieberger,
  ``Open \& Closed vs Pure Open String Disk Amplitudes,''
  arXiv:0907.2211 [hep-th].
  %%CITATION = ARXIV:0907.2211;%%
}
\lref\stieseis{
  D.~Oprisa and S.~Stieberger,
  ``Six gluon open superstring disk amplitude, multiple hypergeometric  series
  and Euler-Zagier sums,''
  arXiv:hep-th/0509042.
  %%CITATION = HEP-TH/0509042;%%
}
\lref\richards{
  D.~M.~Richards,
  ``The One-Loop Five-Graviton Amplitude and the Effective Action,''
  JHEP {\bf 0810}, 042 (2008)
  [arXiv:0807.2421 [hep-th]].
  %%CITATION = JHEPA,0810,042;%%
\semi
  D.~M.~Richards,
  ``The One-Loop $H^2R^3$ and $H^2(DH)^2R$ Terms in the Effective Action,''
  JHEP {\bf 0810}, 043 (2008)
  [arXiv:0807.3453 [hep-th]].
  %%CITATION = JHEPA,0810,043;%%
}
\lref\bedoya{
  O.~A.~Bedoya and N.~Berkovits,
  ``GGI Lectures on the Pure Spinor Formalism of the Superstring,''
  arXiv:0910.2254 [hep-th].
  %%CITATION = ARXIV:0910.2254;%%
}
\lref\explaining{
  N.~Berkovits,
  ``Explaining the Pure Spinor Formalism for the Superstring,''
  JHEP {\bf 0801}, 065 (2008)
  [arXiv:0712.0324 [hep-th]].
  %%CITATION = JHEPA,0801,065;%%
}
\lref\fiveone{
  C.~R.~Mafra and C.~Stahn,
  ``The One-loop Open Superstring Massless Five-point Amplitude with the
  Non-Minimal Pure Spinor Formalism,''
  JHEP {\bf 0903}, 126 (2009)
  [arXiv:0902.1539 [hep-th]].
  %%CITATION = JHEPA,0903,126;%%
}
\lref\fivetree{
  C.~R.~Mafra,
  ``Simplifying the Tree-level Superstring Massless Five-point Amplitude,''
  JHEP {\bf 1001}, 007 (2010)
  [arXiv:0909.5206 [hep-th]].
  %%CITATION = JHEPA,1001,007;%%
}
\lref\verlinde{
  E.~P.~Verlinde and H.~L.~Verlinde,
  ``Chiral bosonization, determinants and the string partition function,''
  Nucl.\ Phys.\  B {\bf 288}, 357 (1987).
  %%CITATION = NUPHA,B288,357;%%
}
\lref\Greencoef{
  M.~B.~Green, J.~G.~Russo and P.~Vanhove,
  ``Low energy expansion of the four-particle genus-one amplitude in type II
  superstring theory,''
  JHEP {\bf 0802}, 020 (2008)
  [arXiv:0801.0322 [hep-th]].
  %%CITATION = JHEPA,0802,020;%%
}
\lref\ulf{
  U.~Gran,
  ``GAMMA: A Mathematica package for performing Gamma-matrix algebra and  Fierz
  transformations in arbitrary dimensions,''
  arXiv:hep-th/0105086.
  %%CITATION = HEP-TH/0105086;%%
}
\lref\Basu{
  A.~Basu,
  ``The D**10 R**4 term in type IIB string theory,''
  Phys.\ Lett.\  B {\bf 648}, 378 (2007)
  [arXiv:hep-th/0610335].
  %%CITATION = PHLTA,B648,378;%%
}
\lref\harris{
Griffiths and Harris, 
``Principles of Algebraic Geometry'', [Wiley Classics Library Edition
Published 1994]
}
\lref\dhokerA{
	E.~D'Hoker and D.~H.~Phong,
  	``Two-Loop Superstrings V: Gauge Slice Independence of the N-Point
  	Function,''
  	Nucl.\ Phys.\  B {\bf 715}, 91 (2005)
 	 [arXiv:hep-th/0501196].
  	%%CITATION = NUPHA,B715,91;%%
	E.~D'Hoker and D.~H.~Phong,
  	``Two-Loop Superstrings IV, The Cosmological Constant and Modular Forms,''
  	Nucl.\ Phys.\  B {\bf 639}, 129 (2002)
  	[arXiv:hep-th/0111040].
  	%%CITATION = NUPHA,B639,129;%%
  	E.~D'Hoker and D.~H.~Phong,
  	``Two-Loop Superstrings III, Slice Independence and Absence of Ambiguities,''
  	Nucl.\ Phys.\  B {\bf 636}, 61 (2002)
  	[arXiv:hep-th/0111016].
  	%%CITATION = NUPHA,B636,61;%%
  	E.~D'Hoker and D.~H.~Phong,
  	``Two-Loop Superstrings II, The Chiral Measure on Moduli Space,''
  	Nucl.\ Phys.\  B {\bf 636}, 3 (2002)
  	[arXiv:hep-th/0110283].
  	%%CITATION = NUPHA,B636,3;%%
  	E.~D'Hoker and D.~H.~Phong,
  	``Two-Loop Superstrings I, Main Formulas,''
  	Phys.\ Lett.\  B {\bf 529}, 241 (2002)
  	[arXiv:hep-th/0110247].
  	%%CITATION = PHLTA,B529,241;%%
}
\lref\DhokerLec{
  E.~D'Hoker and D.~H.~Phong,
  ``Lectures on Two-Loop Superstrings,''
  arXiv:hep-th/0211111.
  %%CITATION = HEP-TH/0211111;%%
}
\lref\MorozovReview{
  A.~Morozov,
  ``NSR Superstring Measures Revisited,''
  JHEP {\bf 0805}, 086 (2008)
  [arXiv:0804.3167 [hep-th]].
  %%CITATION = JHEPA,0805,086;%%
}
\lref\tsimpis{
	G.~Policastro and D.~Tsimpis,
        ``R**4, purified,''
        arXiv:hep-th/0603165.
        %%CITATION = HEP-TH 0603165;%%
}
\lref\ictp{
  N.~Berkovits,
  ``ICTP lectures on covariant quantization of the superstring,''
  arXiv:hep-th/0209059.
  %%CITATION = HEP-TH/0209059;%%
}
\lref\morozov{
  A.~A.~Belavin, V.~Knizhnik, A.~Morozov and A.~Perelomov,
  ``Two- and Three-loop Amplitudes in the Bosonic String Theory,''
  JETP Lett.\  {\bf 43}, 411 (1986)
  [Phys.\ Lett.\  B {\bf 177}, 324 (1986)].
  %%CITATION = PHLTA,B177,324;%%
}
\lref\twovan{
  M.~B.~Green, H.~h.~Kwon and P.~Vanhove,
  ``Two loops in eleven dimensions,''
  Phys.\ Rev.\  D {\bf 61}, 104010 (2000)
  [arXiv:hep-th/9910055].
  %%CITATION = PHRVA,D61,104010;%%
}
\lref\theorems{
  N.~Berkovits,
  ``New higher-derivative R**4 theorems,''
  Phys.\ Rev.\ Lett.\  {\bf 98}, 211601 (2007)
  [arXiv:hep-th/0609006].
  %%CITATION = PRLTA,98,211601;%%
}

\newsec{Introduction}

Scattering amplitudes led to the discovery of string theory more than 40 years ago. But
after all these years, explicit results for higher-loop and/or higher-point amplitudes
are relatively sparse. In fact, since the publication of the famous review by D'Hoker
and Phong \dhokerperturbation\ in 1988, there has been a small number of new ten-dimensional 
scattering computations. Using either the RNS or GS formalisms, the extensions to our knowledge 
in higher loops \dhokerVI\ or higher points \refs{\MedinaUm,\stieseis,\stieOC,\richards} were 
limited to bosonic external states while the overall coefficients were not
always under consideration\foot{There are however powerful approaches to discuss the
coefficients which do not require direct ten-dimensional scattering computations \Basu\Greencoef.}.

Since the discovery of the manifestly space-time supersymmetric
pure spinor formalism \refs{\superpoincare,\multiloop,\nonmin,\explaining}
there has been progress in extending results of scattering amplitudes\foot{The 
use of the pure spinor formalism however is not
limited to scattering amplitudes only. For reviews, see \refs{\ictp,\bedoya}.}
to the whole supermultiplet \refs{\multiloop,\twoloop,\twolooptwo,\tsimpis,\anomalia,
\fiveone,\fivetree,\mafraids}
by using the pure spinor superspace \refs{\PSSuperspace} but explicit
computations for genus higher than two are still missing though \refs{\nekrasov,\yuri,\grassi}.
And the amplitudes in the pure spinor formalism were also computed up to 
the overall coefficients. That has changed since \humberto, where the
precise normalizations for the pure spinor measures were determined and where 
it was also shown how to evaluate integrals in pure spinor space.

So in this paper we use and extend the results of \humberto\ to obtain the coefficient
of the type IIB (and IIA \twovan\theorems) two-loop massless four-point amplitude from a first principles
computation and for the whole supermultiplet. To achieve that we use
pure spinor measures which present the feature of having simple forms for all genera,
in deep contrast with the complicated superstring measure for the RNS 
formalism \refs{\DhokerLec,\MorozovReview}.
As mentioned in \dhokerS, it is still an unsolved problem to find the precise
normalizations for the chiral bosonization formul{\ae} of \verlinde.
Therefore the two-loop coefficient can not be obtained from a direct
calculation in the RNS formalism. In fact, computing the amplitude up
to the overall coefficient already required several years of effort which
resulted in an impressive series of papers \refs{\dhokerA,\dhokerVI}, so the strategy
adopted in \dhokerS\ was to {\it fix} the two-loop coefficient
indirectly by using factorization. So in this respect the
calculations of this paper make it very clear how the pure
spinor formalism can surpass the RNS limitations. But to present our
results we have chosen to adopt the clear conventions of \dhokerS,
which also eases the detection of any mismatches.

In section 2 the conventions and several pure spinor specific results 
are written down. Emphasis is made regarding the generality and simplicity
of the pure spinor setup. The computations of the
three- and four-point amplitudes at tree-level are performed in section 3 
to show that the conventions of section 2 match the RNS ones of \dhokerS\ such that
${\cal A}_0^{\rm PS} = {\cal A}_0^{\rm RNS}$, where
$$
{\cal A}^{\rm PS}_0 = (2\pi)^{10}\d^{(10)}(k){\k}^4 e^{-2\l}
\({\sqrt{2} \over 2^{12} \pi^6 \ap^5}\)  \({\a'\over 2}\)^8 K{\bar K}
C(s,t,u)
$$
Then we use the very same machinery of the tree-level computation
to obtain also the full supersymmetric one- and two-loop amplitudes --- including
their precise coefficients --- in sections 4 and 5,
\eqn\umcorreto{
{\cal A}^{\rm PS}_1 = (2\pi)^{10}\d^{(10)}(k) {\k^{4} K{\bar K} \over 2^{9} \pi^2 \ap^5}
\({\ap\over 2}\)^{8} \int_{{\cal M}_1} {d^{2}\tau \over \tau_2^5}
\prod_{i=2}^4 \int d^2 z_i \prod_{i<j}^4 F_1(z_i,z_j)^{\a k^i\cdot k^j},
}
\eqn\doiscorreto{
{\cal A}_2^{\rm PS} =
(2 \pi)^{10}\d^{(10)}(k)\kappa^4 e^{2\l} {\sqrt{2} K {\bar K} \over 2^{10}\ap^5}
 \({\ap\over 2}\)^{10}
\int_{{\cal M}_2} {d^2\Omega_{IJ} \over ({\rm det}{\rm Im}\Omega_{IJ})^5} 
\int_{\Sigma_4} |{\cal Y}_s|^2
\prod_{i < j} F_2(z_i, z_j)^{\a k^i\cdot k^j}
}
which explicitly shows that with the pure spinor formalism
those coefficients follow directly from a first principles computation.
But we find disagreement with the RNS results reported by \dhokerS,
namely
\eqn\mismatches{
{\cal A}_1^{\rm PS} = {1\over 2^2}{\cal A}_1^{\rm RNS}, \quad
{\cal A}_2^{\rm PS} = {1\over 2^4}{\cal A}_2^{\rm RNS}.
}
The mismatches seen in \mismatches\ will deserve some consideration.
On one hand, the previous PS computation of the one-loop coefficient in
\humberto\  by one
of the authors claimed agreement with the RNS result of \dhokerS. But as will be
pointed out in section 4, \humberto\ made a mistake in the evaluation of the
b-ghost integral which explains the difference with the computation of this paper. 
On the other (RNS) hand, we 
argue in section 4 that \dhokerS\ forgot the two factors of $1/2$ from the GSO projection
in the left- and right-moving sectors in their measure.
This observation will also explain the $1/2^4$ mismatch
at two-loops of section 5, as \dhokerS\ fixed the two-loop coefficient using a factorization
constraint which depends quadratically on the one-loop coefficient\foot{For a compact Riemann
surface $S$
of genus $g$ the correct factor is $1/2^{2g}$, which is the number of spin structures over $S$ and
is in agreement with factorization.}.

In the appendix A we present the detailed covariant computation of the 
two-loop kinematic factor needed in section 5. This appendix can be regarded as
a fully $SO(10)$-covariant proof of the 2-loop equivalence\foot{As will be mentioned in appendix A,
there is a loophole in the 2-loop equivalence proof of \anomalia. Some terms in the non-minimal
pure spinor kinematic factor were argued to vanish using a $U(5)$ decomposition
but, as will be shown explicitly using the
identities of \mafraids, are in fact proportional to the kinematic factor of the minimal
pure spinor formalism. As this loophole only affects the proportionality constant, it
does not alter the conclusions of \anomalia\ but had to be taken into account here.}
between the non-minimal 
and minimal pure spinor formalisms, and is analogous to the covariant proof of \teseM\
for the 1-loop case. 
The appendix B is devoted
to proving a formula mentioned {\it en passant} in \twoloop\ which is
used to rewrite the two-loop amplitude in terms
of integrals in the period matrix instead of in the Teichm\"uller parameters.

% Na los!

\newsec{The conventions}

The non-minimal pure spinor formalism action for the left-moving sector reads \nonmin\
\eqn\action{
S = {1\over 2\pi \ap}\int_{\Sigma_g} d^2z \( \p X^m \pb X_m + \ap p_\a \pb \t^\a
- \ap \omega_\a \pb\l^\a
- \ap \wb^\a \pb\lb_\a + \ap s^\a\pb r_\a \)
}
with the constraints $(\l\g^m\l)=(\lb\g^m \lb) = (\lb\g^m r) =0$. The space-time
dimensions are the following \humberto
\eqn\dim{
[\ap] = 2,\; [X^m] = 1, \; [\t^\a] = [\l^\a] = [\bar \omega^\a] = [s^\a] = 1/2, \;
[p_\a]=[\omega_\a]=[\lb_\a]=[r_\a] = -1/2.
}
The OPE's for the matter variables following from \action\ can be computed to be
\eqn\OPEs{
X^m(z)X_n(w) \sim -{\ap\over 2} \d^m_n {\rm ln}|z-w|^2, \quad
p_\a(z) \t^\b(w) \sim {\d^\b_a\over z-w}.
}
The Green-Schwarz constraint $d_\a(z)$ and the supersymmetric momentum $\Pi^m(z)$ are
\eqn\dalpha{
d_\a = p_\a - {1\over \ap}(\g^m\t)_\a \p X_m - {1\over 4\ap}(\g^m \t)_\a(\t \g_m \p\t), \quad
\Pi^m = \p X^m + \half (\t\g^m \p\t)
}
which satisfy the following OPE's
$$
 d_\a (z)d_\b(w) \sim - {2\over \ap}{\g^m_{\a\b}\Pi_m \over z-w},\quad
d_\a(z)\Pi^m(w) \sim {\g^{m}_{\a\b}\p\t^\b \over z-w},
$$
\eqn\opedp{
d_\a(z)f(\t(w),x(w))\sim {D_{\a}f(\theta(w),x(w)) \over z-w}, \quad
\Pi^m(z)f(\t(w),x(w))\sim -{\ap\over 2}{k^m f(\theta(w),x(w)) \over z-w}
}
where 
$D_\a = {\p\over \p\t^\a} + \half (\g^m\t)_\a \p_m$
is supersymmetric derivative. The composite b-ghost is given by \nonmin\ (see also \bOda)
$$
b = s^\a \p\lb_\a + 
{1\over 4(\l\lb)}\( 2\Pi^{m}(\lb \g_{m}d) -
N_{mn}(\lb \g^{mn}\p\t) - J_{\lambda}(\lb \p\t) - (\lb \p^2 \t) \)
$$
$$
+ {(\lb\g^{mnp}r)\over 192 (\l\lb)^2}\big[ {\ap\over 2}(d\g_{mnp}d) + 24N_{mn}\Pi_{p}\big]
- {\ap\over 2}{(r\g_{mnp}r) \over 16(\l\lb)^3 }\big[
(\lb\g^{m}d)N^{np}
- {(\lb\g^{pqr}r)N^{mn}N_{qr} \over 8(\l\lb)}\big],
$$
and satisfies \nonmin\
\eqn\QbT{
\{Q,b(z)\} = T(z)
}
where the BRST-charge $Q$ and the energy-momentum tensor $T(z)$ are
$$
Q=\oint (\l^\a d_\a + \wb^{\a}r_{\a}),\quad
T(z)=-{1\over \ap}\p X^m \p X_m -
p_\a \p\t^\a + \omega_{\a}\p\l^{\a} +
\wb^\a \p\lb_{\a} -s^\a\p r_{\a}.
$$
From \dim\ it follows that $[Q]=[b]=[T] = 0$.

Scattering amplitudes in the non-minimal pure spinor formalism use vertex
operators in unintegrated and integrated forms, which for the massless
states are given respectively by
\eqn\vertices{
V(z) = \l^\a A_\a, \quad U(z) = \p\t^\a A_\a + A_m \Pi^m + {\ap\over 2}d_\a W^\a + {\ap \over 4}N_{mn}{\cal F}^{mn}
}
where $A_\a(X,\t)$, $A^m(X,\t)$, $W^\a(X,\t)$, ${\cal F}^{mn}$
are the standard 10-dimensional ${\cal N} = 1$ SYM superfields \ref\SYM{
	E.~Witten,
        ``Twistor - Like Transform In Ten-Dimensions,''
        Nucl.\ Phys.\ B {\bf 266}, 245 (1986).
        %%CITATION = NUPHA,B266,245;%%
\semi
	W.~Siegel,
        ``Superfields In Higher Dimensional Space-Time,''
        Phys.\ Lett.\ B {\bf 80}, 220 (1979).
        %%CITATION = PHLTA,B80,220;%%
}. They have the following
$\t$-expansion \ref\thetaSYM{
  	J.~P.~Harnad and S.~Shnider,
	``Constraints And Field Equations For Ten-Dimensional Superyang-Mills
  	Theory,''
  	Commun.\ Math.\ Phys.\  {\bf 106}, 183 (1986).
  	%%CITATION = CMPHA,106,183;%%
\semi
	H.~Ooguri, J.~Rahmfeld, H.~Robins and J.~Tannenhauser,
        ``Holography in superspace,''
        JHEP {\bf 0007}, 045 (2000)
        [arXiv:hep-th/0007104].
        %%CITATION = HEP-TH 0007104;%%
\semi
	P.~A.~Grassi and L.~Tamassia,
        ``Vertex operators for closed superstrings,''
        JHEP {\bf 0407}, 071 (2004)
        [arXiv:hep-th/0405072].
        %%CITATION = HEP-TH 0405072;%%
}\tsimpis\
$$
A_{\a}(x,\t)={1\over 2}a_m(\g^m\t)_\a -{1\over 3}(\xi\g_m\t)(\g^m\t)_\a
-{1\over 32}F_{mn}(\g_p\t)_\a (\t\g^{mnp}\t) + \ldots
$$
$$
A_{m}(x,\t) = a_m - (\xi\g_m\t) - {1\over 8}(\t\g_m\g^{pq}\t)F_{pq}
         + {1\over 12}(\t\g_m\g^{pq}\t)(\p_p\xi\g_q\t) + \ldots
$$
$$
W^{\a}(x,\t) = \xi^{\a} - {1\over 4}(\g^{mn}\t)^{\a} F_{mn}
           + {1\over 4}(\g^{mn}\t)^{\a}(\p_m\xi\g_n\t)
	   + {1\over 48}(\g^{mn}\t)^{\a}(\t\g_n\g^{pq}\t)\p_m F_{pq} 
	   + \ldots
$$
$$
\cF_{mn}(x,\t) = F_{mn} - 2(\p_{[m}\xi\g_{n]}\t) + {1\over
4}(\t\g_{[m}\g^{pq}\t)\p_{n]}F_{pq} + {\ldots},
$$
where $a_m(x) = e_m {\rm e}^{ik\cdot x}$, $\xi^{\a}(x) = (2/\ap)^{1/2}\chi^\a {\rm e}^{ik\cdot x}$ and
$F_{mn} = 2\p_{[m} a_{n]}$ with $[e_m] = 0$ and $[\chi^\a]= 1/2$. The space-time dimensions of the superfields
and the vertex operators are
\eqn\Susydim{
[A_\a] = 1/2, \quad [A_m] = 0, \quad [W^\a] = -1/2, \quad [{\cal F}_{mn}] = -1, \quad [V(z)] = [U(z)] = 1.
}
Vertex operators for the closed string are $V(z,\bar z) = {\tilde \kappa} V(z)\otimes {\tilde V}(\bar z)$ and
$U(z,\bar z) = {\tilde \kappa} U(z)\otimes {\tilde U}(\bar z)$ with the understanding that only the left-moving
modes carry the ${\rm e}^{ik\cdot x}$ factor. ${\tilde \kappa}$ is the overall vertex operator 
normalization which will be fixed below to ${\tilde \kappa} = \kappa$, where $\kappa$ is the
normalization convention used in \dhokerS.
Therefore as in \dhokerS, its precise value in terms of $\ap$ and the string coupling constant \weinbergU\
will not be needed here. 

Finally, the string coupling constant appearing in
scattering amplitude computations in the pure spinor formalism is ${\rm e}^{(2g-2)\mu}$. As
discussed below, by choosing a convenient normalization for the pure spinor 
tree-level measures its
equality with the RNS convention of \dhokerS\ ${\rm e}^{(2g-2)\mu} = {\rm e}^{(2g-2)\lambda}$
will follow.

The construction of the zero-mode measures for the non-minimal pure spinor variables was given 
in \nonmin\ and their precise normalizations were found in \humberto. 
It is however convenient to use slightly different conventions for the measures of \humberto\ to
make their genus-$g$ dependence (and generality) explicit, facilitating computations in
different genera. The space-time dimensionless genus-$g$ zero-mode measures are given
by
\eqn\dl{
[d\l] T_{\a_1\a_2\a_3\a_4\a_5}
= c_{\l} \e_{\a_1 ...\a_5 \rho_1 ... \rho_{11}} d\l^{\rho_1} ... d\l^{\rho_{11}}
}
\eqn\dlb{
[d\lb] {\bar T}^{\a_1\a_2\a_3\a_4\a_5}
= c_{\lb} \e^{\a_1 ...\a_5 \rho_1 ... \rho_{11}} d\lb_{\rho_1} ... d\lb_{\rho_{11}}
}
\eqn\dw{
[d\omega] 
= c_{\omega} T_{\a_1\a_2\a_3\a_4\a_5}
\e^{\a_1 ...\a_5 \rho_1 ... \rho_{11}} d\omega_{\rho_1} ... d\omega_{\rho_{11}}
}
\eqn\dwb{
[d\wb] T_{\a_1\a_2\a_3\a_4\a_5}
= c_{\wb} \e_{\a_1 ...\a_5 \rho_1 ... \rho_{11}} d\wb^{\rho_1} ... d\wb^{\rho_{11}}
}
\eqn\dr{
[dr] = c_r {\bar T}^{\a_1\a_2\a_3\a_4\a_5} \e_{\a_1 ...\a_5 \d_1 ... \d_{11}}
\p_{r}^{\d_1} ... \p_{r}^{\d_{11}}
}
\eqn\ds{
[ds^I] = c_s T_{\a_1\a_2\a_3\a_4\a_5}
\epsilon^{\a_1{\ldots} \a_5\rho_1{\ldots} \rho_{11}}
\p^{s^I}_{\rho_1}{\ldots} \p^{s^I}_{\rho_{11}}
}
\eqn\dtmeas{
[d\t] = c_\t d^{16}\t, \quad [dd^I] = c_d d^{16}d^I
}
with the following normalizations 
\eqn\normau{
c_{\l} = \({\ap\over 2}\)^{-2} {1 \over 11!} \({A_g \over 4\pi^2}\)^{11/2}\quad
c_{\omega} = \({\ap\over 2}\)^2{(4\pi^2)^{-11/2}\over 11!5!\, Z_g^{11/g}} 
}
\eqn\normat{
c_{\lb} = \({\ap\over 2}\)^{2} {2^6 \over 11!} \({A_g \over 4\pi^2}\)^{11/2}\quad
c_{\wb} = \({\ap\over 2}\)^{-2} {(4\pi^2)^{-11/2} (\l\lb)^3 \over 11!\, Z_g^{11/g}}
}
\eqn\norma{
c_r = \({\ap\over 2}\)^{-2} {R \over 11!5!}\({2\pi \over A_g}\)^{11/2}\quad
c_s = \({\ap\over 2}\)^2{(2\pi)^{11/2} R^{-1} \over 2^6 11!5! (\l\lb)^3} Z_g^{11/g} 
}
\eqn\normad{
c_{\t} = \({\ap\over 2}\)^{4} \({2\pi\over A_g}\)^{16/2}\quad
c_{d} = \({\ap\over 2}\)^{-4} (2\pi)^{16/2} Z_g^{16/g} 
}
where $R$ is arbitrary and parametrizes the freedom in choosing
the normalization of the tree-level amplitude and $A_g$ is the area of
the Riemann surface. As will be shown in section 3, using the value
\eqn\rnsconv{
R^2= {\sqrt{2}\over 2^{16}\pi}
}
fixes the tree-level normalization to be the same
as in the RNS computations of \dhokerS.
The tensors $T_{\a_1{\ldots} \a_5}$, ${\bar T}^{\a_1{\ldots} \a_5}$ are defined as
\eqn\T{
T_{\a_1\a_2\a_3\a_4\a_5} = (\l \g^m)_{\a_1}(\l \g^n)_{\a_2}(\l \g^p)_{\a_3} (\g_{mnp})_{\a_4\a_5}
}
\eqn\Tbarra{
{\bar T}^{\a_1\a_2\a_3\a_4\a_5} = (\lb \g^m)^{\a_1}(\lb \g^n)^{\a_2}(\lb \g^p)^{\a_3} (\g_{mnp})^{\a_4\a_5}
}
and satisfy
\eqn\Teq{
T_{\a_1\a_2\a_3\a_4\a_5} {\bar T}^{\a_1\a_2\a_3\a_4\a_5} = 5!\, 2^6 (\l\lb)^3.
}
The appearance of the area $A_g$ and of the factor $Z_g$ will be explained in the
next subsection. They are
\eqn\Zg{
A_g = \int d^2z \sqrt{g}, \quad Z_g = {1\over \sqrt{{\rm det}(2{\rm Im}(\Omega_{IJ}))}}, \quad g \ge 1
}
where $\Omega_{IJ}$ is the period matrix of the Riemann surface. It is well-known that for $g=1$ the
period matrix is given by the Teichm\"uller parameter $\tau$.

%\noindent
To avoid cluttering in the formul{\ae} we define the genus $g$ bracket
$\langle \,\rangle_{(n,g)}$ as
\eqn\save{
\langle M(\l,\lb,\t)\rangle_{(n,g)} \equiv 
\int [d\t][dr][d\l][d\lb] {e^{-(\l\lb)-(r\t)}\over (\l\lb)^{3-n}} M(\l,\lb,\t,r)
}
for an arbitrary pure spinor superfield $M(\l,\lb,\t,r)$.
With the above conventions the integral over the zero modes of 
pure spinor space becomes \humberto\
\eqn\humbps{
\int [d\l][d\lb] (\l\lb)^n {\rm e}^{-(\l\lb)} = {(7+n)!\over 7!\, 60}\({2\pi \over A_g}\)^{11},
\quad n \geq 0
}
which together with \Teq\ imply that
\eqn\Ndef{
N_{(n,g)} \equiv \langle \l^3\t^5 \rangle_{(n,g)} 
 = 2^7 R\({ 2\pi\over A_g}\)^{5/2} \({\ap\over 2}\)^2 {(7+n)!\over 7!}, \quad n \geq 0,
}
where we used the abbreviated notation 
$(\l^3\t^5) = (\l\g^r \t)(\l\g^s \t)(\l\g^t \t)(\t\g_{rst}\t)$.
Due to the identities of \mafraids\ the following trick from \humberto\ is required for the tree-level, one- and
two-loop amplitudes
\eqn\humtrick{
\langle(\l A^1)(\l\g^m W^2)(\l\g^n W^3){\cal F}^4_{mn}\rangle_{(n,g)} =
-{K\over 2^9\, 3^2\, 5}\langle (\l^3\t^5)\rangle_{(n,g)}
}
where $K$ denotes the kinematic factor of \dhokerS, which will be written down below.

It is convenient to consider
the genus-g expectation value of the exponentials at the same time as the 
integration over the non-zero modes of the pure spinor variables, as the latter
is equal to $({\rm det}\p\pb)^5$ \humberto. When both expressions
are computed the determinant factors cancels out and one can use the following expression
\eqn\exptree{
\langle \prod_{i=1}^4 e^{ik\cdot x}\rangle_g = (2\pi)^{10}\d^{(10)}(k) {A_g^5 \over (2\pi^2\ap)^5}
\prod_{i < j} F_g(z_i, z_j)^{\a k^i\cdot k^j}
}
for their combined result. Therefore by using \exptree\
the integration over non-zero modes of the pure spinor variables is already taken care of.
For the sphere one has
$F_0(z_i, z_j) = |z_{ij}|$
whereas for genus $g\geq 1$ it can be written in terms of the prime form as \dhokerperturbation
\eqn\primeform{
F_g(z_i, z_j)^{\a k^i\cdot k^j} = |E(z_i,z_j)|^{\a k^i\cdot k^j}{\rm exp}(-2\pi({\rm
Im}\Omega)^{-1}_{IJ}({\rm Im}\int_{z_i}^{z_j}w_I)({\rm Im}\int_{z_i}^{z_j}w_J) ),
}
where $w_I(z)$ ($I=1,...,g$) are the holomorphic 1-forms over $\Sigma_g$.

From \Ndef\ and \exptree\ it follows that in amplitudes of closed string states the factors
of $A_g$ cancel in the always-present product of,
\eqn\simpleA{
|N_{(n,g)}|^2 \langle \prod_{i=1}^N e^{ik\cdot x}\rangle_g
= (2\pi)^{10}\d^{(10)}(k) {\sqrt{2} \over 2^2 \pi^6\ap^5}
\({\ap\over 2}\)^{4} \({(7+n)!\over 7!}\)^2 \prod_{i < j} F_g(z_i, z_j)^{\a k^i\cdot k^j}.
}
The independence of the closed string amplitude with respect to the area of the surface 
follows from the fact that the number of bosonic and fermionic conformal weight-zero variables
is the same.

The topological prescription \nonmin\ for computing the 4-point amplitudes at tree-level,
one- and two-loops\foot{The $\half$ factor appearing in the two-loop 
amplitude was argued for in \bershadsky. Every Riemann surface of genus 2 can be written like a
hyperelliptic curve $y^2=h(z)$
where $h(z)$ is a polynomial of degree 6 and $y$ is the coordinate over $CP^1$. This curve has
the $Z_2$ symmetry $y\,\rightarrow\, -y$, so the 1/2 factor is needed. We would like to thank Cumrun
Vafa for this explanation.} is
\eqn\treelevel{
{\cal A}_0 = {\tilde \k}^{4}e^{-2\mu}\int d^{2}z_4
\langle |{\cal N} \, V^1(0)V^2(1)V^3(\infty)U^4(z_4)|^2 \rangle
}
\eqn\oneloop{
{\cal A}_1 = {1\over 2}{\tilde \k}^{4} \int_{{\cal M}_1} d^{2}\tau_1
\prod_{i=2}^4 \int d^2 z_i \langle |{\cal N} \, (b,\mu_1) V^1(0) U^i(z_i)|^2 \rangle
}
\eqn\twoloops{
{\cal A}_2 = \half {\tilde \k}^4 e^{2\mu} \int_{{\cal M}_2} \prod_{I=1}^3 d^2\tau_I \prod_{i=1}^4
\int d^2z_i
\langle |{\cal N} (b,\mu_I) U^i(z_i)|^2\rangle
}
where ${\cal M}_1$ (${\cal M}_2$) is the fundamental domain of the Riemann surface of genus 1 (genus
2) and ${\cal N}$ is the regulator \nonmin\
\eqn\regulator{
{\cal N} = \sum_{I=1}^g {\rm e}^{-(\l\lb) - (w^I\wb^I) - (r\t) + (s^Id^I)}
}
$\langle\;\rangle$ denotes the integrations over the zero-modes
\eqn\zeromodes{
\langle \;\rangle \rightarrow \prod_{I=1}^g\int [d\theta][dd^I][dr][ds^I][d\wb^I][dw^I][d\l][d\lb]
}
and the b-ghost insertion is \refs{\polchinski,\PolchinskiFactor}
\eqn\binsert{
(b,\mu_j) = {1\over 2\pi}\int d^2 y_j b_{zz}\mu_{j\,\bar z}^z, \quad j=1,{\ldots} ,3g-3.
}
where the normalization $1/2\pi$ comes from bosonic string theory \polchinski\  because the
topological prescription is based on it. 
With the above conventions, the space-time dimension of the genus-$g$ four-point amplitudes 
is given by $[{\cal A}_g] = 8$. In the following sections we don't keep track of the 
overall sign of the amplitudes.

Following \dhokerS\ we use $d^2\tau = d\tau \wedge d{\bar \tau}$, $d^2z = dz\wedge d{\bar z}$ (in particular
$\int_{\Sigma_1} d^2z = 2\tau_2$). Furthermore ${\cal Y}_s$ has space-time dimension $-2$ and is given by
\eqn\Ys{
{\cal Y}_s = -s \Delta(1,4)\Delta(2,3) + t\Delta(1,2)\Delta(3,4),
}
where $\Delta(i,j) \equiv w_1(z_i)w_2(z_j) - w_1(z_j)w_2(z_i)$ and $w_I(z)$ is the basis
of holomorphic 1-forms discussed below and 
$s = - 2(k^1\cdot k^2)$,
$t = - 2(k^2\cdot k^3)$,
$u = - 2(k^1\cdot k^3)$ are the Mandelstam variables satisfying $s+t+u = 0$. Finally, the
omnipresent supersymmetric kinematic factor $K$ can be conveniently represented 
by the pure spinor superspace expression 
$K= 23040\langle (\l A^1)(\l\g^m W^2)(\l\g^n W^3){\cal F}^4_{mn}\rangle$, where the brackets
here are defined such that $\langle (\l^3\t^5) \rangle = 1$ \mafraids. While the computations of \dhokerS\
did not involve the whole supermultiplet, this representation of $K$ is convenient 
because its bosonic component expansion
has the same normalization of the kinematic factor $K$ of \dhokerS,
\eqn\toito{
K = (e^1\cdot e^2)\big[2tu(e^3\cdot e^4) - 4t(k^1\cdot e^3)(k^2\cdot e^4)\big] + {\rm perm} + {\rm fermions}
}
where the fermionic terms can be looked up in \mafraids.

%********************************************
\subsec{The normalization of zero-modes}
%********************************************

Since the dimension of the zero \v{C}ech cohomology
group $H^{0}(\Sigma_g,\Omega^{1})$, where $\Omega^{1}(\Sigma_g)$ is the sheaf of holomorphic 1-forms
over $\Sigma_g$, is equal to the genus {\it g} of the Riemann surface we
expand a generic conformal weight (1,0) field as \nonmin\
\eqn\expansion{
\phi(z) = {\hat \phi}(z) + \sum_{i=1}^g w_i(z)\phi^i
}
where %$\phi$ represents each component of the fields $(w_{\a},\wb^{\a},s^{\a},d_{\a})$,
$\phi^i$ are the zero modes and $\{w_i(z)dz\}$ is a basis of the
$H^{0}(\Sigma_g,\Omega^{1})$ group such that
$$
\int_{a_i}w_j(z)dz=\d_{ij}, \qquad
\int_{b_i}w_j(z)dz=\Omega_{ij}\qquad  i,j=1,2,...,g
$$
\eqn\product{
(w_i,w_j) \equiv \int_{\Sigma_g}w_i\,\bar w_j\,\, dz\wedge d\bar z  = 2 \rm{Im} \Omega_{ij}
}
where $a_i$ and $b_j$ are the generators of the $H^1(\Sigma_{g},Z)=Z^{2g}$
homology group and $\Omega_{ij}$ is the period matrix \harris.
If we expand $\phi$ over another basis $\{\a_j\}$
related by $w_i = B_i^j \a_j$ then \Beilinson,
$$
{\rm det}\(\int_{\Sigma_g}\, w_i\, \bar w_j\,dz\wedge d\bar z\)
= {\rm det}|B|^2{\rm det}\(
\int_{\Sigma_g}\a_i\,\bar \a_j\,\, dz\wedge d\bar z\)
$$
so that for 
\eqn\jacobian{
|{\rm det}B| = \sqrt{{\rm det}(2{\rm Im}\Omega_{ij})} = Z_g^{-1}
}
the basis $\{\a_j\}$ is orthonormal, $(\a_i,\a_j) = \d_{ij}$.
Expanding the fields over the new basis as
$\phi= \sum_{j=1}^g \phi^{\prime j}\a_j$
one can show that the measure satisfies 
\eqn\measures{
d\phi^{\prime 1}\cdots d\phi^{\prime g} 
= {\rm det(B)}^\e d\phi^{1}\cdots d\phi^{g},
}
where $\e = + 1 (-1)$ for bosonic (fermionic) fields.
In the non-minimal formalism the integration measures for conformal
weight-one fields is defined in terms of the $\phi^\prime$ components, but
it is more convenient to use the $\{w_I\}$ basis in explicit computations.
To account for this we absorb the Jacobian \jacobian\ equally into each
of the $[d\phi^I]$ measures as $({\rm det(B)}^{\e/g} d\phi^{1})
\cdots ({\rm det(B)}^{\e/g} d\phi^{g})$, which 
explains the factors of $Z_g$ in \normau\ -- \normad.

Similarly, the appearance of $A_g$ in the measures of the conformal weight-zero 
variables $[\l^\a, \lb_\a, r_\a, \t^\a]$ 
follows from the expansion in a complete set of eigenfunctions for
the Laplacian of the worldsheet \Weisberger
\eqn\zeromo{
\l^\a(z) = \l^\a_0 \Lambda_0 + \sum_{j} \l^\a_j \Lambda_j(z,{\bar z})
}
and $\Lambda_0= 1$ is the generator of the cohomology group $H^{0}(\Sigma_{g},\cal{O})=\bf C$,
where $\cal{O}$ is the sheaf of holomorphic functions over $\Sigma_g$. Because the norm of
$\Lambda_0$ is
$||\Lambda_0||^2 = A_g$
the measures of the scalars must have the Jacobian $A_g ^{\e/2}$ (where $\e = + 1 (-1)$ for
bosonic (fermionic) fields), explaining the factors of $A_g$ in \normau\ -- \normad.

%The factors of $A_g$ in the measures therefore account
%for the fact that we integrate over the zero-modes (e.g. $\l^\a_0$) but the fields in the
%integrand are written with the $\Lambda_0$ factors absorbed  (e.g. $\l^\a \equiv
%\l^\a_0\Lambda_0$).

%*******************************************************
\subsec{On the normalization of the holomorphic 1-forms}
%*******************************************************

The result of scattering amplitudes in the pure spinor formalism does not depend
on the normalization of the holomorphic 1-forms $w_I(z)$.
To see this one
notes that in closed string amplitudes\foot{The analysis can be
trivially modified to the open string.} at genus $g$ the difference between
the number of independent fermionic and bosonic 
conformal weight-one left-moving variables is always $16g + 11g - 11g - 11g = 5g$, corresponding
to $d_\a^I, s^{\a\, I}, w^I_\a$ and $\wb^{\a\, I}$. As $Z_g$ appear in the conformal
weight-one measures as $Z_g^{1/g}$, their total contribution to closed string amplitudes
is always $|Z_g^5|^2 = Z_g^{10}$. Furthermore, when saturating the $11g$ $s^{\a\, I}$ zero modes
the regulator factor ${\cal N}$ provides $11g$ $d^I_\a$ zero-modes as well -- because they
appear in the combination $(s^Id^I)$ in ${\cal N}$ and there is nowhere else to get
$s^{I\, \a}$ zero-modes from. So to complete the saturation of $d_\a^I$
the b-ghosts and external vertices will always provide $5g$ factors of 
$|d_\a^I w_I(z)|^2$, which scales as $x^{10g}$ under $w_I(z) \rightarrow x w_I(z)$.
To finish the proof it suffices to note from \product\ and \jacobian\ that
$Z_g$ scales as $Z_g \rightarrow x^{-g} Z_g$ and therefore $|Z_g^{5}|^2$ offsets
the scaling of the $|w_I^{5g}|^2$ factors from the b-ghosts and external vertices.

%*********************************
\newsec{Tree-level}
%*********************************

The  massless four-point amplitude at tree-level is given by \treelevel,
\eqn\fourpres{
{\cal A}_0 = {\tilde \k}^{4}e^{-2\mu}\int d^{2}z_4
\langle |{\cal N} \, V^1(0)V^2(1)V^3(\infty)U^4(z_4)|^2 \rangle.
}
The amplitude \fourpres\ was computed in components by \tsimpis\ and later expressed in pure spinor superspace
up to an overall normalization in  \mafraids, where it was used that
$\langle \prod_{i=1}^4 e^{i k^i x(z_i,{\bar z}_i)} \rangle = |z_4|^{-\half \ap t} |1 - z_4|^{-\half \ap u}$.
The normalization of the tree-level amplitude of \mafraids\ can be determined {\it a posteriori}
by using the precise value for the expectation value of the exponentials,
\eqn\corrigir{
\langle \prod_{i=1}^4 e^{i k^i x(z_i,{\bar z}_i)} \rangle_0 = (2\pi)^{10}\d^{(10)}(k)
\({A_0\over 2\pi^2 \ap}\)^5
|z_4|^{-\half \ap t} |1 - z_4|^{-\half \ap u},
}
where $A_0 = 4\pi$ is the area of the sphere.
Doing that in the computations of \mafraids\ we obtain,
\eqn\respq{
{\cal A}_0 =  (2\pi)^{10}\d^{(10)}(k){\tilde \k}^4 e^{-2\mu}
\({4\pi\over 2\pi^2\ap}\)^5  \({\a'\over 2}\)^4 K_0{\bar K}_0
C(s,t,u),
}
where 
\eqn\short{
C(s,t,u) =  2\pi {\Gamma(-{\displaystyle \a' s\over 4})\Gamma(-{\displaystyle \a' t\over 4})
\Gamma(-{\displaystyle \a' u\over 4}) 
\over \Gamma(1+{\displaystyle \a' s\over 4})\Gamma(1+{\displaystyle \a' t\over 4})
\Gamma(1+{\displaystyle \a' u\over 4})}
}
and the kinematic factor $K_0$ is given by the pure spinor superspace expression \mafraids\
\eqn\treefour{
K_0 = \langle (\l A^1)(\l\g^m W^2)(\l\g^n W^3){\cal F}^4_{mn}\rangle_{(3,0)}
= -{K\over 2^9\, 3^2\, 5}\langle (\l^3\t^5)\rangle_{(3,0)}
}
where the last equality follows from \humtrick. Using \Ndef\ we get 
\eqn\comp{
K_0 = K {N^{(3,0)} \over (2^9\,3^2\,5)}
= {R\over \sqrt{2}} \({\ap \over 2}\)^2 K,
}
and therefore
\eqn\qptresp{
{\cal A}_0 =  (2\pi)^{10}\d^{(10)}(k){\tilde \k}^4 e^{-2\mu}
{R^2\over 2}\({2 \over \pi\ap}\)^5  \({\a'\over 2}\)^8 K{\bar K}
C(s,t,u)
}
$$
= (2\pi)^{10}\d^{(10)}(k){\tilde \k}^4 e^{-2\mu}
\({\sqrt{2} \over 2^{12} \pi^6 \ap^5}\)  \({\a'\over 2}\)^8 K{\bar K}
C(s,t,u),
$$
where we used that $R^2 = {\sqrt{2}\over 2^{16}\pi}$.

%***********************************************************************
\subsec{The tree-level normalization}
%***********************************************************************

To fix the normalizations at tree-level to match those of \dhokerS\ we need
two conditions \weinbergU, therefore we also evaluate the three-point amplitude,
which is given by
\eqn\tpt{
{\cal A}_{\rm t} = {\tilde \k}^{3}e^{-2\mu} \langle |{\cal N} V(0)V(1)V(\infty)|^2 \rangle.
}
Using \exptree, the component expansion found in \teseM\ and the fact that $(k^i\cdot k^j) = 0$
$$
{\cal A}_{\rm t} = (2\pi)^{10}\d^{(10)}(k) 
{\tilde \k}^3 e^{-2\mu} {A_0^5 \over (2\pi^2\ap)^5} |K_t|^2
$$
hence,
\eqn\tptresp{
{\cal A}_{\rm t} = (2\pi)^{10}\d^{(10)}(k) {\tilde \k}^3 e^{-2\mu}  {\sqrt{2} \over 2^6 \pi^6 \ap^5}
\({\ap\over 2}\)^4 W_3 {\bar W}_3
}
where we used that
$$
|K_t|^2 = |\langle (\l A^1)(\l A^2)(\l A^3)\rangle_{(3,0)}|^2
=  {|N_{(3,0)}|^2\over 2880^2} W_3{\bar W}_3 = {\sqrt{2}\over 2^6\pi}
\({2\pi\over A_0}\)^5\({\ap\over 2}\)^4 W_3 {\bar W}_3
$$
and $W_3 = (e^1\cdot e^2)(k^2\cdot e^3) + (e^1\cdot e^3)(k^1\cdot e^2) + (e^2\cdot e^3)(k^3\cdot e^1)$
is the 3-pt kinematic factor in the RNS computation of \dhokerS. 

In the normalization conventions of \dhokerS\ the tree-level tree- and four-point amplitudes were shown
to be given by\foot{Note that $[{\cal A}_t] = 6$ and $[{\cal A}_0] = 8$, so in \dhokerS\
the factors of $(\ap/2)$ were forgotten.}
\eqn\tDhoker{
{\cal A}^{\rm RNS}_{\rm t} = (2\pi)^{10}\d^{(10)}(k)
\k^3 e^{-2\l} \({\sqrt{2}\over 2^6 \pi^6 \ap^5}\) \({\ap\over 2}\)^4 W_3 {\bar W}_3,
}
\eqn\qDhoker{
{\cal A}^{\rm RNS}_0 =  (2\pi)^{10}\d^{(10)}(k)\k^4 e^{-2\l}
   \({\sqrt{2} \over 2^{12} \pi^6\ap^5}\) \({\a'\over 2}\)^8 K{\bar K}
C(s,t,u).
}
Comparing the RNS results of \tDhoker\ and \qDhoker\ with the corresponding PS 
amplitudes of \tptresp\ and \qptresp\ it follows that
\eqn\solnorm{
{\tilde \k} = \k, \quad e^{-2\mu} = e^{-2\l},
}
so the PS and RNS tree-level normalization conventions are the same.
The numerical value of the parameter $R$ in \rnsconv\ was chosen
precisely for this match to happen. After this tree-level matching is done there remains no more
freedom to adjust conventions.

%*********************************
\newsec{One-loop}
%*********************************

The one-loop massless four-point amplitude is given by \oneloop,
\eqn\fourone{
{\cal A}_1 = {1\over 2}\k^{4} \int_{{\cal M}_1} d^{2}\tau_1
\prod_{i=2}^4 \int d^2 z_i \langle |{\cal N} \, (b,\mu_1) V^1(0) U^i(z_i)|^2 \rangle.
}
The regulator in \regulator\ becomes ${\cal N} = {\rm e}^{-(\l\lb) - (w^1\wb^1) - (r\t) + (s^1d^1)}$,
$\langle\quad\rangle$ denotes the  integrations over the
zero-modes of 
$[\t^\a, d_\a, r_\a, s^\a, w_\a, \wb^\a, \l^\a, \lb_\a]$ and the b-ghost
insertion written in \binsert\ reads
\eqn\bfant{
(b,\mu_1) = {1\over 2\pi}\int d^2 z b_{zz}\mu_{\bar z}^z.
}
As discussed in \nonmin, there is an unique way to saturate the zero-modes of all
variables. The b-ghost must provide two $d^1_\a$ zero-modes with
${1\over 2^6 3}({\ap\over 2})(\lb\g^{mnp}r)(d^1\g_{mnp}d^1)w_1w_1$,
where $w_1 = 1$ is the holomorphic 1-form in the torus.
Therefore the integral \bfant\ is easily computed to give
$$
(b,\mu_1) = 
{1\over 2^7 3\pi}\({\ap\over 2}\){(\lb \g^{mnp}r)(d^1\g_{mnp}d^1)\over (\l\lb)^2},
$$
because $\int d^2 z w_1w_1\mu_1 = 1$. 
The integrated vertices contribute three $d^1_\a$ zero-modes via
$({\ap\over 2})^3 (d^1W^2)(d^1W^3)(d^1W^4)$, so \fourone\ becomes
\eqn\interm{
{\cal A}_1 = {1\over 2^{15}3^2\,\pi^2} \k^{4} \({\ap\over 2}\)^8 
\int_{{\cal M}_1} d^{2}\tau \prod_{i=2}^4 \int d^2 z_i
|{\cal K}_1|^2 \langle\prod_{i=1}^4 {\rm e}^{ikX(z_i)}\rangle_1,
}
where the computation of the zero-mode integrations in 
$$
{\cal K}_1 = \int [dd^1][ds^1][dw^1][d\wb^1]
\,{\rm e}^{- (w^1\wb^1) + (s^1d^1)} \times
$$
\eqn\Kzero{
\times \langle (\lb \g^{mnp}r)(d^1\g_{mnp}d^1) (\l A^1) (d^1W^2)(d^1W^3)(d^1W^4)\rangle_{(1,1)}
}
is straightforward and goes as follows. Using the measures \dw\ and \dwb\
and the results of \humberto\ one gets
\eqn\wint{
\int [dw][d\wb] {\rm e}^{-(w\wb)} = {(\l\lb)^3\over (2\pi)^{11}Z_1^{22}}.
}
Hence,
\eqn\Kzerod{
{\cal K}_1 ={1\over (2\pi)^{11}Z_1^{22}} \int [dd^1][ds^1] {\rm e}^{(s^1d^1)}
\langle (\lb \g^{mnp}r)(d\g_{mnp}d)(\l A^1) (dW^2)(dW^3)(dW^4)\rangle_{(4,1)}.
}
The integration over $[ds]$ using the measure \ds\ leads to
$$
{\cal K}_1 ={(2\pi)^{-11/2}\over 2^{6}(11!5!) Z_1^{11} R}\({\ap\over 2}\)^2
\int [dd^1] T_{\a_1{\ldots} \a_5}\e^{\a_1{\ldots} \a_5 \d_1{\ldots} \d_{11}}
d_{\d_1}{\ldots} d_{\d_{11}}
$$
\eqn\Kzerot{
\langle (\lb \g^{mnp}r) (d^1\g_{mnp}d^1)(\l A^1) (d^1W^2)(d^1W^3)(d^1W^4)\rangle_{(1,1)}.
}
Using the identities
\eqn\epsinta{
\int d^{16}d d_{\rho_1}{\ldots} d_{\rho_{16}} =
\epsilon_{\rho_1{\ldots} \rho_{16}}, \quad \quad
\epsilon_{\rho_1{\ldots} \rho_{16}}\epsilon^{\a_1{\ldots} \a_5\rho_1{\ldots} \rho_{11}}=
11!5!\d^{\a_1{\ldots} \a_5}_{\rho_{12}{\ldots} \rho_{16}},
}
\eqn\noveseisa{
(\g^{abc})^{\rho_{12}\rho_{13}}(\g_{m_1n_1p_1})_{\rho_{12}\rho_{13}} = -2^5 3 \,\d^{abc}_{m_1n_1p_1},
}
\eqn\antissima{
(\l\g^{m_1})_{[\a_1} (\l\g^{n_1})_{\a_2}(\l\g^{p_1})_{\a_3}(\g_{m_1n_1p_1})_{\a_4\a_5]} =
(\l\g^{m_1})_{\a_1} (\l\g^{n_1})_{\a_2}(\l\g^{p_1})_{\a_3}(\g_{m_1n_1p_1})_{\a_4\a_5}
}
the integration over $[dd^1]$ is easily performed and
\Kzerot\ becomes
\eqn\Kzeroq{
{\cal K}_1 ={3(2\pi)^{5/2} Z_1^5 \over 2 R} \({\ap\over 2}\)^{-2}
\langle (\lb \g^{mnp}D) (\l A^1) (\l\g_m W^2)(\l\g_n W^3)(\l\g_p W^4)
\rangle_{(1,1)}
}
where we also used that \anomalia\ 
$\int {\rm e}^{-(r\t)}r_\a ({\ldots} ) = \int D_\a {\rm e}^{-(r\t)} ({\ldots} )$.
Using the identity \teseM
$$
\langle (\lb \g^{mnp}D) (\l A^1) (\l\g_m W^2)(\l\g_n W^3)(\l\g_p W^4)\rangle_{(1,1)} 
= 40\langle (\l A^1) (\l\g^m W^2)(\l\g^n W^3){\cal F}^4_{mn}\rangle_{(2,1)}
$$
$$
= {K\over 2^6\, 3^2}\langle (\l^3\t^5) \rangle_{(2,1)}
$$
where in the last line we used \humtrick, the kinematic factor \Kzeroq\ 
can be written as
\eqn\Kzeroc{
{\cal K}_1 ={(2\pi)^{5/2} Z_1^{5} K \over 3R \,2^{7}}\({\ap\over 2}\)^{-2}
\langle (\l^3\t^5) \rangle_{(2,1)}
}
Using the definition \Ndef\ one concludes from \Kzeroc\ that
\eqn\squared{
|\langle {\cal K}_1 \rangle|^2 = {(2\pi)^5 Z_1^{10} \over 2^{14} 3^2R^2}
K{\bar K}|N_{(2,1)}|^2\({\ap\over 2}\)^{-4}.
}
The amplitude \interm\ therefore is given by
$$
{\cal A}_1 = {(2\pi)^{5}\over 2^{29} 3^4R^2\pi^2} K{\bar K}\k^{4} \({\ap\over 2}\)^{4}
\int_{{\cal M}_1} d^{2}\tau Z_1^{10} \prod_{i=2}^4 \int d^2 z_i
|N_{(2,1)}|^2\langle\prod_{i=1}^4 {\rm e}^{ikX(z_i)}\rangle_1
$$
which upon using \simpleA,
$$
|N_{(2,1)}|^2\langle\prod_{i=1}^4 {\rm e}^{ikX(z_i)}\rangle = 
(2\pi)^{10}\d^{(10)}(k) {2^{25}3^4 R^2\over (2\pi)^5\ap^5}\({\ap\over 2}\)^4\prod_{i < j} 
F_1(z_i, z_j)^{\a k^i\cdot k^j}
$$
and $Z_1^{10} = (2\tau_2)^{-5}$ finally becomes
\eqn\resultOne{
{\cal A}_1 = (2\pi)^{10}\d^{(10)}(k) {\k^{4} K{\bar K} \over 2^{9} \pi^2 \ap^5}
\({\ap\over 2}\)^{8} \int_{{\cal M}_1} {d^{2}\tau \over \tau_2^5}
\prod_{i=2}^4 \int d^2 z_i \prod_{i<j}^4 F_1(z_i,z_j)^{\a k^i\cdot k^j}.
}
It should be pointed out that the previous computation in \humberto\ claimed that 
the 1-loop computation in the pure spinor formalism
agreed with the RNS result of \dhokerS, but it was incorrectly
used that $\int d^2z w_1 w_1 \mu_{\bar z}^z = 2$ instead of $=1$.
And to compare with the result of \dhokerS\ one takes
into account the translation invariance of the torus to integrate the ``extra''
$\int {d^2 z_1\over \tau_2} = 2$ integral in their equation (2.22) to conclude that
\resultOne\ differs\foot{There is a missing factor of $(\ap/2)^8$ in \dhokerS.}
by ${1\over 4}$ from the RNS result reported in \dhokerS. We argue that
the one-loop result of \dhokerS\ is missing the two factors of $1/2$
from the GSO projection for both the left- and right-moving sectors,
explaining the $1/2^2$ discrepancy\foot{We thank Eric D'Hoker
for kindly confirming to us their missing $1/4$ factor \ref\email{Eric D'Hoker,
private communication.}.}.
   
%***********************************************************************
\newsec{Two-loop}
%***********************************************************************

The two-loop massless four-point amplitude in the non-minimal pure spinor formalism
is given by
\eqn\amplitude{
{\cal A}_2 = \half 
\k^4 e^{2\l} \prod_{i=1}^4 \prod_{j=1}^3\int_{{\cal M}_2} d^2\tau_j \int d^2z_i
\langle |{\cal N}  (b,\mu_j) U^i(z_i)|^2\rangle
}
where $\langle \rangle$ denote the zero-mode integrations
$\prod_{I=1}^2\int [d\theta][dd^I][dr][ds^I][d\wb^I][dw^I][d\l][d\lb]$ and
\eqn\bgb{
(b,\mu_j) = {1\over 2\pi}\int d^2 y_j b_{zz}\mu_{j\,\bar z}^z.
}
The 32 (22) zero-modes of
$d_\a$ ($s^\a$) are denoted by $d^I_\a$ ($s^\a_I$) for $I=1,2$. As shown
in \nonmin, they 
are saturated by the different factors of \amplitude\ as
\eqn\explain{
{\cal N} \rightarrow (s^1 d^1)^{11}(s^2 d^2)^{11} \quad \prod_{j=1}^3 (b,\mu_j) \rightarrow (d^1)^3 (d^2)^3
\quad U^1U^2U^3U^4 \rightarrow (d^1)^2 (d^2)^2,
}
so that each b-ghost contributes only zero-modes with the term $({\ap\over 2}){(\lb \g^{mnp} r)\over
192 (\l\lb)^{2}}(d\g_{mnp}d)$. 
The expansion $d_\a(y_i) = {\hat d}_\a(z) + d^1_\a w_1(y_i) + d^2_\a w_2(y_i)$ implies a zero-mode
contribution of
$$
(d\g_{mnp}d)(y) = (d^1\g_{mnp}d^1)f_{11}(y) + 2 (d^1\g_{mnp}d^2)f_{12}(y)
+ (d^2\g_{mnp}d^2)f_{22}(y)
$$
where $f_{ij}(y) \equiv w_i(y)w_j(y)$, $i,j = 1,2$ is the basis of holomorphic quadratic
differentials for the genus-2 Riemann surface \farkas.
It follows from a short computation that,
$$
\prod_{j=1}^3 (b,\mu_j)
= c_b  \prod_{j=1}^3 \int d^2 y_j \mu_j(y_j)
\Delta(y_1,y_2)\Delta(y_2,y_3)\Delta(y_3,y_1)
$$
\eqn\bghostD{
{1\over (\l\lb)^{6}}(\lb\g_{abc}r)(\lb\g_{def}r)(\lb\g_{ghi}r)
(d^1 \g^{abc} d^1)(d^1 \g^{def} d^2)(d^2 \g^{ghi} d^2)
}
where $c_b = {2 \over (384\pi)^3}({\ap\over 2})^3$ and $\Delta(y,z)
= w_1(y)w_2(z) - w_2(y)w_1(z)$.
In the computation of \bghostD\ one can check that combinations
containing a different number of $d^1_\a$ and $d^2_\a$ zero modes e.g.,
$$
(\lb\g_{abc}r)(\lb\g_{def}r)(\lb\g_{ghi}r)
(d^1 \g^{abc} d^2)(d^1 \g^{def} d^2)(d^2 \g^{ghi} d^2)
$$
vanish trivially due to the index symmetries, confirming the zero mode
counting of \explain.
Using the period matrix parametrization of moduli space the b-ghost
insertions become
$$
\int_{{\cal M}_2} d^2\tau_1d^2\tau_2 d^2\tau_3 |\prod_{j=1}^3 (b,\mu_j)|^2
= 
$$
$$
= c_b^2 \int_{{\cal M}_2} d^2 \Omega_{IJ}
|{1\over (\l\lb)^{6}}(\lb\g_{abc}r)(\lb\g_{def}r)(\lb\g_{ghi}r)
(d^1 \g^{abc} d^1)(d^1 \g^{def} d^2)(d^2 \g^{ghi} d^2)|^2
$$
where $\int d^2 \Omega_{IJ} = \int d^2\Omega_{11}d^2\Omega_{12}d^2\Omega_{22}$ and
we used the identity of the appendix B.

The integration over $[dw^I][d\wb^I]$ can be done using the results of \humberto\ taking
into account the different normalizations for the measures \dw\ and \dwb,
\eqn\dablios{
\int [dw^1][d\wb^1][dw^2][d\wb^2]e^{-(w^1\wb^1) -(w^2\wb^2)} = {(\l\lb)^6\over (2\pi)^{22}}Z_2^{-22}
}
It is straightforward to use the measure \ds\ to integrate over $[ds^1][ds^2]$, and
the amplitude \amplitude\ becomes
$$
{\cal A}_2 = { \kappa^4 e^{2\l} \over 2^{56} \pi^{26} 3^6 (11!5!)^4}
\({\ap\over 2}\)^8 \int_{{\cal M}_2} d^2\Omega_{IJ}
|Z_2^{-11}\int [d\t][dd^1][dd^2][dr][d\l][d\lb]
$$
$$
{{\rm e}^{-(\l\lb)-(r\t)} \over (\l\lb)^6}(\lb\g_{abc}r)(\lb\g_{def}r)(\lb\g_{ghi}r)
(d^1 \g^{abc} d^1)(d^1 \g^{def} d^2)(d^2 \g^{ghi} d^2)
$$
$$
(\l\g^{m_1})_{\a_1} (\l\g^{n_1})_{\a_2}(\l\g^{p_1})_{\a_3}(\g_{m_1n_1p_1})_{\a_4\a_5}
(\l\g^{m_2})_{\b_1} (\l\g^{n_2})_{\b_2}(\l\g^{p_2})_{\b_3}(\g_{m_2n_2p_2})_{\b_4\b_5}
$$
$$
\epsilon^{\a_1{\ldots} \a_5\rho_1{\ldots} \rho_{11}}
\epsilon^{\b_1{\ldots} \b_5\d_1{\ldots} \d_{11}}
d^1_{\rho_1}{\ldots} d^1_{\rho_{11}}
d^2_{\d_1}{\ldots} d^2_{\d_{11}}
$$
$$
\big[
(d^1 W^1)(d^1 W^2)(d^2 W^3)(d^2 W^4)w_1(z_1)w_1(z_2)w_2(z_3)w_2(z_4)
$$
$$
+ (d^1 W^1)(d^2 W^2)(d^1 W^3)(d^2 W^4)w_1(z_1)w_2(z_2)w_1(z_3)w_2(z_4)
$$
$$
+ (d^1 W^1)(d^2 W^2)(d^2 W^3)(d^1 W^4)w_1(z_1)w_2(z_2)w_2(z_3)w_1(z_4)
$$
$$
+ (d^2 W^1)(d^2 W^2)(d^1 W^3)(d^1 W^4)w_2(z_1)w_2(z_2)w_1(z_3)w_1(z_4)
$$
$$
+ (d^2 W^1)(d^1 W^2)(d^1 W^3)(d^2 W^4)w_2(z_1)w_1(z_2)w_1(z_3)w_2(z_4)
$$
\eqn\exprds{
+ (d^2 W^1)(d^1 W^2)(d^2 W^3)(d^1 W^4)w_2(z_1)w_1(z_2)w_2(z_3)w_1(z_4)\big]|^2
\times \langle \prod_{i=1}^4 e^{ik\cdot x}\rangle_2
}
where the only non-vanishing contribution 
from the external vertices contains two $d^1$ and two $d^2$ zero-modes coming from
$(\ap/2)^4 (dW)^4$.
Integrating the $d_\a$ zero-modes in \exprds\ using \dtmeas\ and
\epsinta\ --- \antissima\
one gets
\eqn\calcula{
{\cal A}_2 =
{\pi^{6} \over  2^{4}3^2}\left({\ap\over 2}\right)^{6}
\int_{{\cal M}_2} d^2\Omega_{IJ} Z_2^{10}
\big| {\cal K}_2 \big|^2
\times \langle \prod_{i=1}^4 e^{ik\cdot x}\rangle_2
}
where the non-minimal kinematic factor ${\cal K}$ is given by
$$
{\cal K}_2 = \langle
(\lb\g_{m_1n_1p_1}r)(\lb\g_{def}r)(\lb\g_{m_2n_2p_2}r)
(\l\g^{m_1defm_2}\l)\big[
$$
$$
+ (\l\g^{n_1} W^1)(\l\g^{p_1} W^2) (\l\g^{n_2} W^3)(\l\g^{p_2} W^4)
\(H_{1234} +H_{3412}\)
$$
$$
+ (\l\g^{n_1} W^1)(\l\g^{p_1} W^3)(\l\g^{n_2} W^2) (\l\g^{p_2} W^4)
\( H_{1324} + H_{2413}\)
$$
\eqn\grande{
+ (\l\g^{n_1} W^1)(\l\g^{p_1} W^4)(\l\g^{n_2} W^2) (\l\g^{p_2} W^3)
\(H_{1423} + H_{2314} \)\big]\rangle_{(-3,2)}
}
and we defined
\eqn\defH{
H_{ijkl} = w_1(z_i)w_1(z_j)w_2(z_k)w_2(z_l).
}
In the Appendix A we will show that
\eqn\mostrar{
{\cal K}_2  = 2^{12}\, 3^3\, 5 \,{\cal Y}_s 
\langle(\l A^1)(\l\g^m W^2)(\l\g^n W^3){\cal F}^4_{mn}\rangle_{(0,2)}
= 2^3\, 3 \, {\cal Y}_s \, K \langle (\l^3\t^5)\rangle_{(0,2)}
}
where the second equality follows from \humtrick.
Hence \calcula\ is given by
\eqn\amplifi{
{\cal A}_2 =
\kappa^4 e^{2\l} 2^2\pi^{6} K {\bar K}\left({\ap\over 2}\right)^{6}
\int_{{\cal M}_2} d^2\Omega_{IJ}Z_2^{10} |{\cal Y}_s|^2 |N_{(0,2)}|^2\langle \prod_{i=1}^4
e^{ik\cdot x}\rangle_2.
}
From the formula \simpleA\ we get
\eqn\facil{
|N_{(0,2)}|^2\langle \prod_{i=1}^4 e^{ik\cdot x}\rangle_2 
= (2\pi)^{10}\d^{(10)}(k) {\sqrt{2} \over 2^2 \pi^6\ap^5}
\({\ap\over 2}\)^{4}  \prod_{i < j} F_2(z_i, z_j)^{\a k^i\cdot k^j}
}
which together with $Z_2^{10} = 2^{-10} {\rm det}({\rm Im}\Omega_{IJ})^{-5}$ 
implies that
\eqn\resultTwo{
{\cal A}_2 =
(2 \pi)^{10}\d^{(10)}(k)\kappa^4 e^{2\l} {\sqrt{2} K {\bar K} \over 2^{10}\ap^5}
 \({\ap\over 2}\)^{10}
\int_{{\cal M}_2} {d^2\Omega_{IJ} \over ({\rm det}{\rm Im}\Omega_{IJ})^5} 
\int_{\Sigma_4} |{\cal Y}_s|^2
\prod_{i < j} F_2(z_i, z_j)^{\a k^i\cdot k^j}
}
which is the final result for the 2-loop amplitude\foot{The coefficient
obtained here is $1/16$ times the result reported by \dhokerS. This difference
can be accounted for by the missing factor of $1/4$ in their 1-loop result which is
used as input in their fixing of the 2-loop coefficient through factorization.}.
And we have shown that the computation of the whole supersymmetric 
amplitude including its coefficient is straightforward using the non-minimal 
pure spinor formalism.

%******************************************
\newsec{Conclusions}
%******************************************

We used the genus-$g$ measures in the non-minimal pure spinor formalism to find
the overall coefficient of the two-loop amplitude and have shown
that there are no major differences in carrying out the computations when
compared against the analogous calculations for the tree-level and one-loop
amplitudes. In fact, this task is significantly simplified by
the pure spinor superspace identities of \mafraids\ linking the four-point
kinematic factors.
These observations must be compared against the unsolved difficulties
in the RNS formalism, which besides having no explicit computations for
the whole supermultiplet has to rely on a factorization procedure to
find the two-loop coefficient. Furthermore, we argued that the
mismatch of $1/16$ found in the two-loop amplitude compared with the result
of \dhokerS\ is due to a missing factor of $1/4$ from the GSO projection
in their one-loop amplitude.

\vskip 15pt {\bf Acknowledgements:} 
CRM and HG would like to thank Eric
D'Hoker, Nathan Berkovits and Stefan Theisen for discussions.
CRM acknowledges support by the Deutsch-Israelische
Projektkooperation (DIP H52). 
HG  acknowledges support by
FAPESP Ph.D grant 07/54623-8.

%*********************************************************************
\appendix{A}{Non-minimal two-loop kinematic factor}
%*********************************************************************

The non-minimal two-loop computation of section 5 leads to the
kinematic factor
\eqn\first{
K = \langle (\lb \g^{abc} D)(\lb \g^{ghi} D)(\lb \g^{def} D)(\l\g_{adefg}\l)\big[
(\l\g^b W^1)(\l\g^c W^2)(\l\g^h W^3)(\l\g^i W^4)
\big]\rangle_{(-3,2)}.
}
In \anomalia\ it was shown\foot{There is a loophole in the proof
of \anomalia\ though. In that proof the terms in \first\ which are of the
form $kWWWF$ where argued to vanish after summing over the permutations. However
we show here that by using the identities of \mafraids\ 
those terms are actually  proportional to $W{\cal F}{\cal F}{\cal F}$, so
the conclusions of \anomalia\ still hold true. CM would like to acknowledge a 
question made by I.~Park which sparked the motivation to revisit that proof.}
that \first\ is proportional to
$\langle (\l\g^{mnpqr}\l)(\l\g^s W){\cal F}_{mn}{\cal F}_{pq}{\cal F}_{rs}\rangle_{(0,2)}$,
the kinematic factor
obtained in the minimal pure spinor formalism \twoloop, whose equivalence with the
RNS result of \dhokerVI\ was established in \refs{\twolooptwo,\mafraids}.
We will now evaluate all the terms in \first\ to find the exact 
coefficient announced in \mostrar.

To simplify the covariant computation of \first\ we use
$(\lb \g^{def} D)(\l\g_{adefg}\l) = 48(\l\lb)(\l\g^{ag}D)
- 48 (\l\g^{ag}\lb)(\l D)$ and drop the last term because $(\l\g^m W^I)$ is BRST-closed.
And for the same reason we can use $(\l\g^a\g^g D)$ instead of $(\l\g^{ag}D)$ in the first
term. Therefore \first\ becomes
\eqn\sec{
K = 48 \langle (\lb\g^{ghi}D)(\l\g^a\g^g D)(\lb\g^{abc}D)\big[
(\l\g^b W^1)(\l\g^c W^2)(\l\g^h W^3)(\l\g^i W^4)
\big]\rangle_{(-2,2)}.
}
The strategy to evaluate and simplify\foot{These kind of computations
confirm the observations made long ago that pure spinors simplify
the description of super-Yang-Mills theory \bigHowe.} \sec\ is straightforward due to the 
identities obeyed by the pure spinor $\l^\a$.  One uses the
SYM equation of motion for $W^\a$ in the form of 
\eqn\super{
(\lb\g^{abc}D)(\l\g^m W^1) = {1\over 4}(\l\g^m \g^{m_1n_1}\g^{abc}\lb){\cal F}^1_{m_1n_1}
}
\eqn\Shakira{
(\l\g^a\g^g D)(\l\g^m W^2) = {1\over 4}(\l\g^{agm_2n_2m}\l){\cal F}^2_{m_2n_2}
}
and uses gamma matrix identities\foot{The package GAMMA \ulf\ is often very useful for these
manipulations.} in such a way as to get factors which vanish by
the pure spinor property of $(\l\g^m)_\a(\l\g_m)_\b = 0$. For example, one gets identities
like
\eqn\outra{
(\l\g^b \g^{m_1n_1}\g^{abc}\lb)(\l\g^a\g^g D)\big[ {\cal F}^1_{m_1n_1}(\l\g^c W^2)\big]
= 48(\l\lb)(\l\g^a\g^g D)\big[ {\cal F}^1_{ac}(\l\g^c W^2)\big]
}
and 
$$
{\cal F}^3_{rs}(\l\g^h \g^{rs}\g^{abc}\lb)(\l\g^a)_\a(\l\g^b)_\b(\l\g^c)_\g =
$$
\eqn\legal{
16(\l\lb)(\d^h_b {\cal F}^3_{ac} -\d^h_c {\cal F}^3_{ab} - \d^h_a {\cal F}^3_{bc})
(\l\g^a)_\a(\l\g^b)_\b(\l\g^c)_\g.
}
Following the above steps \sec\ becomes
$$
K= 576\langle(\lb\g^{ghi}D)(\l\g^a\g^g D)\big[
{\cal F}^1_{ab}(\l\g^b W^2) (\l\g^h W^3)(\l\g^i W^4)
$$
$$
- {1\over 3}{\cal F}^3_{ab}(\l\g^b W^1)(\l\g^h W^2)(\l\g^i W^4)
- {1\over 3}{\cal F}^4_{ab}(\l\g^b W^1)(\l\g^h W^2)(\l\g^i W^3) + (1 \leftrightarrow 2)
\big]\rangle_{(-1,2)}
$$
\eqn\giga{
-192\langle(\lb\g^{gai}D)(\l\g^a\g^g D)\big[ {\cal F}^3_{bc}(\l\g^i W^4)
(\l\g^b W^1)(\l\g^c W^2) + (3\leftrightarrow 4)\big]\rangle_{(-1,2)}.
}
The last line of \giga\ vanishes. To see this note that the factor inside brackets
is BRST-closed, so that we can replace $(\l\g^a\g^g D)$ by $(\l\g^{ag} D)$. Furthermore
$(\lb\g^{gai}D)(\l\g^{ga}D) = -  (\lb\g^{ga}\g^i D)(\l\g^{ga}D) - 2(\lb \g^a D)(\l\g^{ia}D)$
and the last term vanishes when acting on ${\cal F}^3_{bc}(\l\g^i W^4)
(\l\g^b W^1)(\l\g^c W^2)$ because $(\l\g^{ia}D) = (\l\g^i\g^a D) - \d^i_a (\l D)$ and
$(\l\g^i)_\a(\l\g_i)_\b = 0$ due to the pure spinor property. Therefore by using the
gamma matrix identity of
\eqn\gamid{
(\g^{mn})_{\a}^{{\phantom m} \d}(\g_{mn})_\b^{{\phantom m}\s} = 
-8 \d_\a^\s \d_\b^\d - 2\d_\a^\d \d_\b^\s + 4 \g^m_{\a\b}\g_m^{\d\s}
}
and dropping the term proportional to the BRST charge and using momentum conservation
(so that $D_\a$ and $D_\b$ effectively anti-commute) we get
\eqn\zerodo{
(\lb\g^{ga}\g^i D)(\l\g^{ga}D) = 8 (\l\lb)(D\g^iD) + 4 (\lb\g^m D)(\l\g^m\g^i D).
}
The first term in the RHS of \zerodo\ is proportional to $k^i$ and vanishes
by momentum conservation, while the last term vanishes when acting on
${\cal F}^3_{bc}(\l\g^i W^4)
(\l\g^b W^1)(\l\g^c W^2)$ for the same reason as explained above.

For convenience we write \giga\ as
\eqn\isso{
K = 576 K_{a_1} - 192  K_{a_2} - 192 K_{a_3} + (1\leftrightarrow 2)
}
where
$$
K_{a_1} \equiv \langle (\lb\g^{ghi}D)(\l\g^a\g^g D)\big[
{\cal F}^1_{ab}(\l\g^b W^2) (\l\g^h W^3)(\l\g^i W^4)\big]\rangle_{(-1,2)}
$$
while $K_{a_2}$ and $K_{a_3}$ can be obtained by permuting the labels in $K_{a_1}$.
Using the SYM equations of motion
and a few gamma matrix identities we get
$$
 K_{a_1} = + \langle(\lb \g^{ghi} D)\Big[ 6 k^1_c(\l\g^g W^1)(\l\g^c W^2)(\l\g^h W^3)(\l\g^i W^4)
$$
$$
-{1 \over 4} (\l\g^{mn pq g}\l){\cal F}^1_{mn}{\cal F}^2_{pq}(\l\g^h W^3)(\l\g^i W^4)
-{1 \over 4} (\l\g^{agmn h}\l){\cal F}^1_{ac}{\cal F}^3_{mn}(\l\g^c W^2)(\l\g^i W^4)
$$
\eqn\chato{
-{1 \over 4} (\l\g^{agmn h}\l){\cal F}^1_{ac}{\cal F}^4_{mn}(\l\g^c W^2)(\l\g^i W^3)\Big]\rangle_{(-1,2)}.
}
After a long and tedious computation using straightforward manipulations and identities like
$(\l\g^{mnpqr}\l){\cal F}^I_{mn}{\cal F}^J_{pq} =
(\l\g^{mnpqr}\l){\cal F}^J_{mn}{\cal F}^I_{pq}$ and \twoloop\
\eqn\elimin{
(\l\g^{mnpqr}\l)(\l\g^s W^4)\big[
{\cal F}^1_{mn}{\cal F}^2_{pq}{\cal F}^3_{rs}
+ {\cal F}^3_{mn}{\cal F}^1_{pq}{\cal F}^2_{rs}
+ {\cal F}^2_{mn}{\cal F}^3_{pq}{\cal F}^1_{rs}\big] = 0
}
one gets
$$
K_{a_1} = -{1\over 2} \langle k^1_m (\lb\g^{ghi}\g^n W^1){\cal F}^2_{pq}(\l\g^{mnpqg}\l)(\l\g^h W^3)
(\l\g^i W^4)\rangle_{(-1,2)} + (1\leftrightarrow 2)
$$
$$
-{1\over 4} \langle \( 2 {\cal F}^3_{rs} k^1_{[a} (\lb\g^{ghi}\g_{c]} W^1) 
+ 2 k^3_r (\lb\g^{ghi}\g^s W^3){\cal F}^1_{ac}\) (\l\g^{agrsh}\l)(\l\g^c W^2) (\l\g^i W^4)\rangle_{(-1,2)}
$$
$$
-{1\over 4} \langle \( 2 {\cal F}^4_{rs} k^1_{[a} (\lb\g^{ghi}\g_{c]} W^1) 
+ 2 k^4_r (\lb\g^{ghi}\g^s W^4){\cal F}^1_{ac}\) (\l\g^{agrsh}\l)(\l\g^c W^2) (\l\g^i W^3)\rangle_{(-1,2)}
$$
$$
+ \langle(\l\g^{mnpqr}\l)\big[
\({\cal F}^1_{mn}{\cal F}^3_{pq}{\cal F}^2_{rs} -4 {\cal F}^1_{mn}{\cal F}^2_{pq}{\cal F}^3_{rs}
\) (\l\g^s W^4)  - 3{\cal F}^3_{mn}{\cal F}^4_{pq}{\cal F}^1_{rs}(\l\g^s W^2) + (3\leftrightarrow 4) \big]
$$
$$
-72 k^1_m (\l\g^m W^2)\big[ {\cal F}^1_{hi}(\l\g^h W^3)(\l\g^i W^4)
+ {\cal F}^3_{hi}(\l\g^h W^1)(\l\g^i W^4)
+ {\cal F}^4_{hi}(\l\g^h W^1)(\l\g^i W^3) 
\big]
$$
\eqn\boring{
+24  k^1_m (\l\g^m W^4) {\cal F}^2_{hi}(\l\g^h W^1)(\l\g^i W^3)
+24  k^1_m (\l\g^m W^3) {\cal F}^2_{hi}(\l\g^h W^1)(\l\g^i W^4)\rangle_{(0,2)}
}
To simplify the $\langle\;\rangle_{(-1,2)}$ terms in \boring\ it is 
convenient to have $\lb_\a$ in the combination $(\l\lb)$ 
by using the identities,
\eqn\mago{
(\lb\g^{ghi}\g^n W^1)(\l\g^{mnpqg}\l)(\l\g^h W^3)
(\l\g^i W^4)
= 2 (\l\lb) (W^3\g^{gi}\g_n W^1)(\l\g^{mnpqg}\l)(\l\g^i W^4)
}
and similarly
\eqn\magd{
(\lb\g^{ghi}\g^a W^1)(\l\g^{agrsh}\l)(\l\g^c W^2)
(\l\g^i W^4)
= 2 (\l\lb) (W^4\g^{ahi} W^1)(\l\g^{ahirs}\l)(\l\g^c W^2)
}
\eqn\magt{
(\lb\g^{ghi}\g^c W^1)(\l\g^{agrsh}\l)(\l\g^c W^2)
(\l\g^i W^4)
= 2 (\l\lb) (W^4\g^{ghc} W^1)(\l\g^{agrsh}\l)(\l\g^c W^2).
}
In \mafraids\ it was proved that
\eqn\mariposa{
\langle (\l\g^{mnpqr}\l)(\l\g^s W^4)
{\cal F}^1_{mn}{\cal F}^2_{pq}{\cal F}^3_{rs}\rangle_{(n,g)} = - 16 (k^1\cdot k^2)
\langle (\l A^1)(\l\g^m W^2)(\l\g^n W^3){\cal F}^4_{mn}\rangle_{(n,g)}
}
and that $\langle (\l A^1)(\l\g^m W^2)(\l\g^n W^3){\cal F}^4_{mn}\rangle_{(n,g)}$ is
completely symmetric in the particle labels, hence
$$
\langle(\l\g^{mnpqr}\l)\big[
\({\cal F}^1_{mn}{\cal F}^3_{pq}{\cal F}^2_{rs} -4 {\cal F}^1_{mn}{\cal F}^2_{pq}{\cal F}^3_{rs}
\) (\l\g^s W^4)  - 3{\cal F}^3_{mn}{\cal F}^4_{pq}{\cal F}^1_{rs}(\l\g^s W^2)\big]\rangle_{(0,2)} 
$$
$$
+ (3\leftrightarrow 4)
= + 240 (k^1\cdot k^2)\langle (\l A^1)(\l\g^m W^2)(\l\g^n W^3){\cal F}^4_{mn}\rangle_{(0,2)},
$$
where we also used the momentum conservation relation of
$(k^1\cdot k^3) + (k^1\cdot k^4) = - (k^1\cdot k^2)$.
The last two lines of \boring\ can be simplified by using
$(\l\g^m W) = QA^m - k^m (\l A)$ and by noticing that the terms of the form
$Q(A^m){\cal F}_{pq}(\l\g^p W)(\l \g^q W)$ are BRST exact and therefore
vanish. Doing that one gets
$$
-72 \langle k^1_m (\l\g^m W^2)\big[ {\cal F}^1_{hi}(\l\g^h W^3)(\l\g^i W^4)
+ {\cal F}^3_{hi}(\l\g^h W^1)(\l\g^i W^4)
+ {\cal F}^4_{hi}(\l\g^h W^1)(\l\g^i W^3) 
\big]
$$
$$
+24 k^1_m (\l\g^m W^4) {\cal F}^2_{hi}(\l\g^h W^1)(\l\g^i W^3)
+24 k^1_m (\l\g^m W^3) {\cal F}^2_{hi}(\l\g^h W^1)(\l\g^i W^4)\rangle_{(0,2)}
$$
\eqn\simplesDois{
= + 240 (k^1\cdot k^2)\langle(\l A^1)(\l\g^m W^2)(\l\g^n W^3){\cal F}^4_{mn}\rangle_{(0,2)}.
}

Feeding the results above into the expression for $K_{a_1}$ in \boring\
one can write it as
$K_{a_1} = K_{a_{11}} + K_{a_{12}}$, where
$$
K_{a_{11}} = - \langle k^1_r (\l\g^{mnpqr}\l)(W^3\g_{mns} W^1)(\l\g^s W^4){\cal F}^2_{pq}\rangle_{(0,2)}
+ (1\leftrightarrow 2)
$$
\eqn\Kuu{
- \big[ \langle({\cal F}^3_{rs} k^1_{[a} (W^4\g_{gh|c]} W^1)
+  k^3_r (W^4\g_{ghs} W^3){\cal F}^1_{ac})
(\l\g^{agrsh}\l)  (\l\g^c W^2)\rangle_{(0,2)} + (3\leftrightarrow 4)\big]
}
and
\eqn\Kud{
K_{a_{12}} =
+ 480 (k^1\cdot k^2)\langle (\l A^1)(\l\g^m W^2)(\l\g^n W^3){\cal F}^4_{mn}\rangle_{(0,2)}
}
Furthermore, by using the gamma
matrix identities $\g^{mnp} = \g^{mn}\g^p - \eta^{mn} \g^p + \eta^{am}\g^n$ and
$$
(\g^{mn})_\a^{\;\; \d}(\g_{mn})_\b^{\;\; \s} =
-8 \d_\a^\s \d_\b^\d + 4 \g^m_{\a\b}\g_m^{\d\s} - 2\d_\a^\d\d_\b^\s,
$$
the pure
spinor identities $(\l\g^{amnpq}\l)(\l\g_a)_\b = (\l\g^m)_\a(\l\g_m)_\b = 0$,
the equation of motion $k^I_m (\l\g^m W^I) = 0$ and the results above,
$K_{a_{11}}$ (and its permutations $K_{a_{21}}$ and  $K_{a_{31}}$) can be further
simplified. In fact, one can show that
$$
-\langle k^1_r(\l\g^{mnpqr}\l)(W^3\g_{mns}W^1)(\l\g^s W^4){\cal F}^2_{pq}\rangle_{(0,2)}
$$
$$
= 32 \langle k^1_m (\l\g^m W^4) (\l\g^p W^3)(\l\g^q W^1){\cal F}^2_{pq}\rangle_{(0,2)} + (3\leftrightarrow 4)
$$
\eqn\longaechata{
= -32\( (k^1\cdot k^3) + (k^1\cdot k^4)\)\langle (\l A^1) (\l\g^m W^2)(\l\g^n W^3){\cal F}^4_{mn}\rangle_{(0,2)}.
}
From $\g^{mnp}_{\a\b}\g_{mnp}^{\g\d} = 48(\d_\a^\g\d_\b^\d - \d_\a^\d\d_\b^\g)$ and the
equation of motion for $W^\a_3$ it follows that,
$$
 -  k^3_r (\l\g^{agrsh}\l) (W^4\g_{ghs} W^3){\cal F}^1_{ac} (\l\g^c W^2)
 = 48(k^3\cdot k^4) (\l A^1) (\l\g^m W^2)(\l\g^n W^3){\cal F}^4_{mn}
$$
and
$$
\half {\cal F}^3_{rs}k^1_c(W^4\g_{gha}W^1)(\l\g^{agrsh}\l)(\l\g^c W^2)
= 48(k^1\cdot k^2)(\l A^1) (\l\g^m W^2)(\l\g^n W^3){\cal F}^4_{mn}.
$$
From \longaechata\ one also gets
\eqn\tediumdd{
-\half {\cal F}^3_{rs}k^1_a(W^4\g_{ghc}W^1)(\l\g^{agrsh}\l)(\l\g^c W^2)
= 16(k^1\cdot k^3)(\l A^1) (\l\g^m W^2)(\l\g^n W^3){\cal F}^4_{mn}.
}
Plugging the identities \longaechata\ -- \tediumdd\ in \Kuu\ and summing over the indicated
permutations leads to
\eqn\mygod{
K_{a_{11}} = 240 (k^1\cdot k^2)\langle (\l A^1)(\l\g^m W^2)(\l\g^n W^3){\cal F}^4_{mn}\rangle_{(0,2)}
}
hence
\eqn\somaKu{
K_{a_1} = K_{a_{11}} + K_{a_{12}} = 720 (k^1\cdot k^2)\langle
(\l A^1)(\l\g^m W^2)(\l\g^n W^3){\cal F}^4_{mn}\rangle_{(0,2)}.
}
From \isso\ and \somaKu\ and their permutations
one arrives at the final result\foot{To check results we performed explicit component
expansion computations with especially-crafted programs using FORM \FORM.}
for \first,
$$
K = + 720 \langle (\l A^1)(\l\g^m W^2)(\l\g^n W^3){\cal F}^4_{mn}\rangle_{(0,2)}\times
$$
$$
\times \big[ 576(k^1\cdot k^2) - 192(k^3\cdot k^2) - 192 (k^4\cdot k^1)
+ 576(k^2\cdot k^1) - 192(k^3\cdot k^1) - 192 (k^4\cdot k^2)
\big]
$$
\eqn\Kfinal{
 =  3\cdot 2^7\cdot 2880 (k^1\cdot k^2) \langle (\l A^1)(\l\g^m W^2)(\l\g^n W^3){\cal F}^4_{mn}\rangle_{(0,2)}.
}

The complete kinematic factor \grande\ is obtained using the result \Kfinal\ and permuting
its labels. The first line of \grande\ is given by \Kfinal\ while the second and third are
obtained by replacing
$s\rightarrow u$ and $s\rightarrow t$ respectively.
The final result is therefore
$$
{\cal K}_2 = - 3\cdot 2^6\cdot 2880 \langle (\l A^1)(\l\g^m W^2)(\l\g^n W^3){\cal F}^4_{mn}\rangle_{(0,2)}
\big[
$$
$$
  s (H_{1234} +H_{3412})
+ u (H_{1324} +H_{2413})
+ t (H_{1423} +H_{2314})\big]
$$
\eqn\above{
= 2^{12}\, 3^3 \, 5 \, {\cal Y}_s \langle
(\l A^1)(\l\g^m W^2)(\l\g^n W^3){\cal F}^4_{mn}\rangle_{(0,2)}
}
where we used the Mandelstam variables 
and $u = - t - s$ together with
$$
H_{1234} +H_{3412} - H_{1324} - H_{2413} = \Delta(1,4)\Delta(2,3)
$$
$$
H_{1423} +H_{2314} - H_{1324} - H_{2413} = - \Delta(1,2)\Delta(3,4).
$$
and the definition \Ys.
With \above\ the expression for the kinematic factor \grande\ is finally
demonstrated.

%*********************************************************************
\appendix{B}{Period matrix parametrization of genus-two moduli space}
%*********************************************************************
Let $\mu_{i\,\,\bar z}^{\,\,z}$  ($i=1,2,3$) be the Beltrami differentials, $\tau_i$ ($i=1,2,3$) the
Teichm\"uller parameters and $w_I(z)$ ($I=1,2$) the holomorphic 1- forms over $\Sigma_2$, then \twoloop\
\eqn\Nidentity{
\int d^2 \tau_{1} d^2 \tau_{2} d^2 \tau_{3} \Big| \prod_{i=1}^{3} \int
d^{2}z_{i}\mu_{i}(z_{i})\Delta(1,2)\Delta(2,3)\Delta(3,1)\Big|^2 =\int d^2 \Omega_{11} d^2
\Omega_{12}d^2 \Omega_{22}
}
where $\Delta(i,j)=w_1(z_i)w_2(z_j)-w_1(z_j)w_2(z_i)$. To prove this one uses 
the identity\foot{In the Mathematics literature this is the ``Rauch variational formula'', see e.g.
\ref\livroazul{
I.~Chavel, H.~M.~Farkas, ``Differential Geometry and Complex Analysis: A volume dedicated to 
the memory of Harry Ernest Rauch'', Springer-Verlag Berlin, Heidelberg 1985.
}\ref\livrohannover{
Nag, Subhashis. ``The Complex Analytic Theory of Teichm\"uller Spaces'', John Wiley \& Sons, 1988
}\ref\rauch{
H.~E.~Rauch, ``On the Transcendental Moduli of Algebraic Riemann Surfaces'',
Proceedings of the National Academy of Sciences of the United States of America, Vol. 41, No. 1 (Jan. 15, 1955), pp. 42-49
}
}
\morozov\dhokerperturbation
\eqn\Didentity{
\int d^{2} z\,\, w_{I}\,(z)w_{J}(z)\,\mu_{i}(z)={\delta \Omega_{IJ} \over \delta\tau_{i}}
}
and expands $\Delta(1,2)\Delta(2,3)\Delta(3,1)$ to get
\eqn\easy{
\prod_{i=1}^{3} \int
d^{2}z_{i}\mu_{i}(z_{i})\Delta(1,2)\Delta(2,3)\Delta(3,1)= -{\delta \Omega_{11} \over
\delta\tau_{i}}
{\delta \Omega_{12} \over \delta\tau_{j}}{\delta \Omega_{22} \over \delta\tau_{k}}\epsilon^{ijk}.
}
So
$$
d\tau_{1}\wedge d\tau_2\wedge d\tau_3 \prod_{i=1}^{3} \int
d^{2}z_{i}\mu_{i}(z_{i})\Delta(1,2)\Delta(2,3)\Delta(3,1)= -{\delta \Omega_{11} \over
\delta\tau_{i}}
{\delta \Omega_{12} \over \delta\tau_{j}}{\delta \Omega_{22} \over
\delta\tau_{k}}\epsilon^{ijk}d\tau_{1}\wedge d\tau_2\wedge d\tau_3
$$
$$
=-{\delta \Omega_{11} \over
\delta\tau_{i}}
{\delta \Omega_{12} \over \delta\tau_{j}}{\delta \Omega_{22} \over
\delta\tau_{k}}d\tau_{i}\wedge d\tau_{j}\wedge d\tau_{k}
$$
$$
=-\delta \Omega_{11}\wedge 
\delta \Omega_{12} \wedge \delta \Omega_{22}. 
$$
Multiplying the last expression by its complex conjugate we get \Nidentity.

\listrefs

\end